\documentclass[12pt]{article}
\usepackage{ulem}
\usepackage {graphicx}
\usepackage{amssymb}
\usepackage{amsmath}
\usepackage{amsbsy}
\usepackage{setspace}
\usepackage{natbib}
\usepackage{bm}
\usepackage{textcomp}
\usepackage{color}
\usepackage{booktabs}
\usepackage{threeparttable}
\usepackage{amsthm}
\usepackage{enumitem}
\usepackage{longtable,threeparttablex,booktabs}
\usepackage{hyperref}
\usepackage{mathtools}
\usepackage{lscape}
\usepackage{subfigure}
\usepackage{multirow}
\usepackage{comment}
\usepackage{lineno}
\allowdisplaybreaks[4]
\newtheorem{theorem}{Theorem}
\newtheorem{corollary}{Corollary}
\newtheorem{proposition}{Proposition}
\newtheorem{lemma}{Lemma}
\newtheorem{example}{Example}

\newtheorem{definition}{Definition}

\def\text#1{\mbox{\rm #1}}
\def\overset#1#2{\stackrel{#1}{#2} }

\def\underwiggle 1{
	\ifmmode\setbox\TempBox=\hbox{$ 1$}\else\setbox\TempBox=\hbox{
		1}\fi \setbox\TempBoxA=\hbox to \wd\TempBox{\hss\char'176\hss}
	\rlap{\copy\TempBox}\smash{\lower9pt\hbox{\copy\TempBoxA}} }

\newcommand{\beq}{\begin{equation}}
	\newcommand{\eeq}{\end{equation}}
\newcommand{\beas}{\begin{eqnarray*}}
	\newcommand{\eeas}{\end{eqnarray*}}
\newcommand{\bea}{\begin{eqnarray}}
	\newcommand{\eea}{\end{eqnarray}}
\newcommand{\bei}{\begin{itemize}}
	\newcommand{\eei}{\end{itemize}}
\newcommand{\ben}{\begin{enumerate}}
	\newcommand{\een}{\end{enumerate}}
\newcommand{\bet}{\begin{theorem}}
	\newcommand{\eet}{\end{theorem}}
\newcommand{\bel}{\begin{lemma}}
	\newcommand{\eel}{\end{lemma}}
\newcommand{\bep}{\begin{proposition}}
	\newcommand{\eep}{\end{proposition}}
\newcommand{\bed}{\begin{definition}}
	\newcommand{\eed}{\end{definition}}
\newcommand{\bec}{\begin{corollary}}
	\newcommand{\eec}{\end{corollary}}
\newcommand{\bex}{\begin{example}}
	\newcommand{\eex}{\end{example}}

\newcommand{\argmin}{\mathop{\rm arg\min}}

\begin{document}
\setcounter{page}{1}
	\title{Regression analysis of mixed sparse synchronous and asynchronous longitudinal covariates with varying-coefficient models}
	
	\author{Congmin Liu$^1$, Zhuowei Sun$^1$ and Hongyuan Cao$^2$ }

	\footnotetext[1]{~ School of Mathematics, Jilin University, Changchun 130012, China.}
    \footnotetext[2]{~ Department of Statistics, Florida State University, Tallahassee, FL 32306, U.S.A.}
	
	\date{}
	\maketitle
	\begin{abstract}
	We consider varying-coefficient models for mixed synchronous and asynchronous longitudinal covariates, where asynchronicity refers to the misalignment of longitudinal measurement times within an individual. We propose three different methods of parameter estimation and inference. The first method is a one-step approach that estimates non-parametric regression functions for synchronous and asynchronous longitudinal covariates simultaneously. The second method is a two-step approach in which synchronous longitudinal covariates are regressed with the longitudinal response by centering the synchronous longitudinal covariates first and, in the second step, the residuals from the first step are regressed with asynchronous longitudinal covariates. The third method is the same as the second method except that in the first step, we omit the asynchronous longitudinal covariate and include a non-parametric intercept in the regression analysis of synchronous longitudinal covariates and the longitudinal response.  
We further construct simultaneous confidence bands for the non-parametric regression functions to quantify the overall magnitude of variation. Extensive simulation studies provide numerical support for the theoretical findings. The practical utility of the methods is illustrated on a dataset from the ADNI study. 
	\end{abstract}
	

	\section{Introduction}\label{intro.sec}
\begin{figure}[!ht]
		\centering
		\subfigure{\includegraphics[width=15cm,height=7cm]{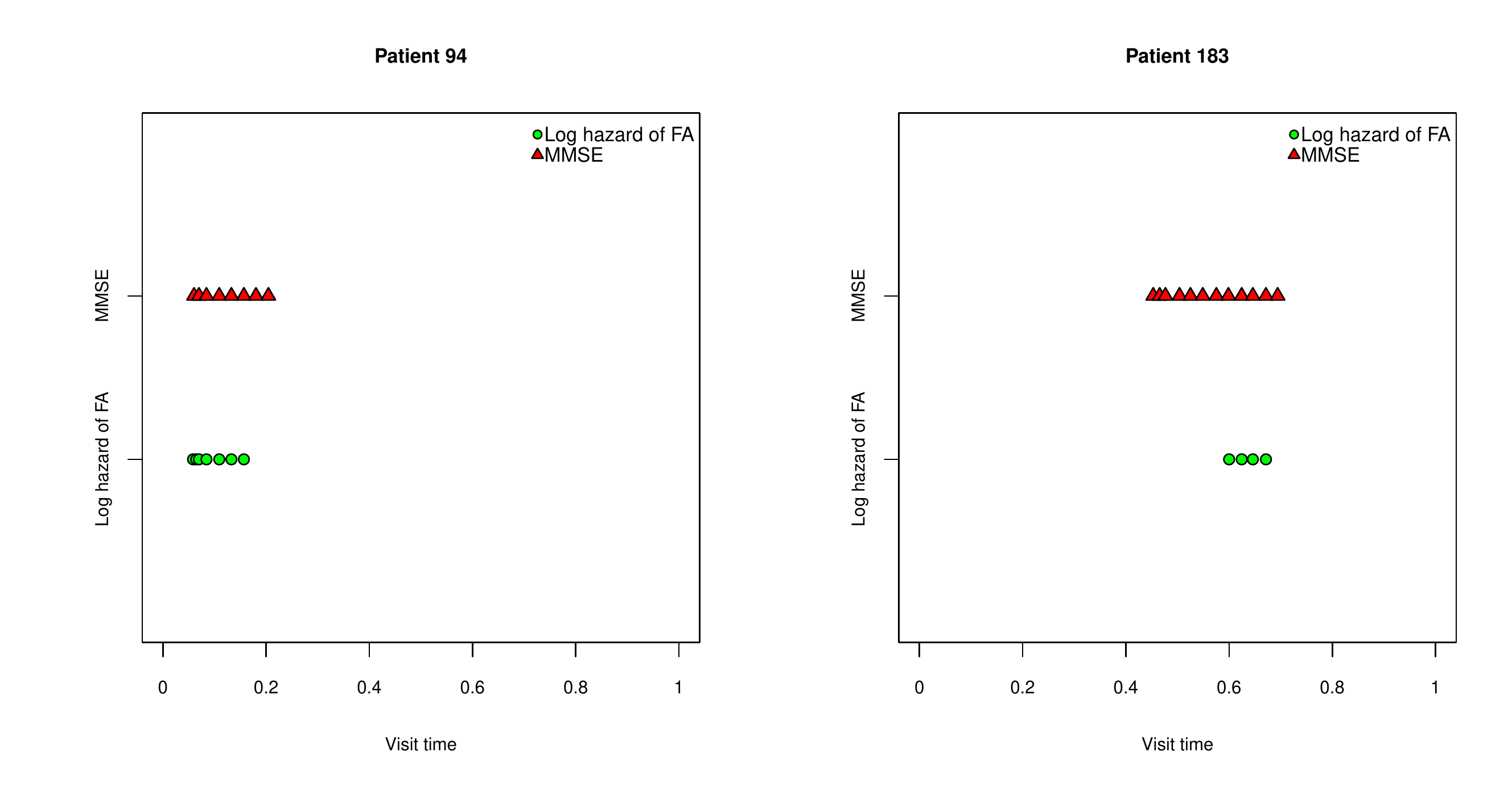}
		}
  
		\subfigure{\includegraphics[width=15cm,height=7.5cm]{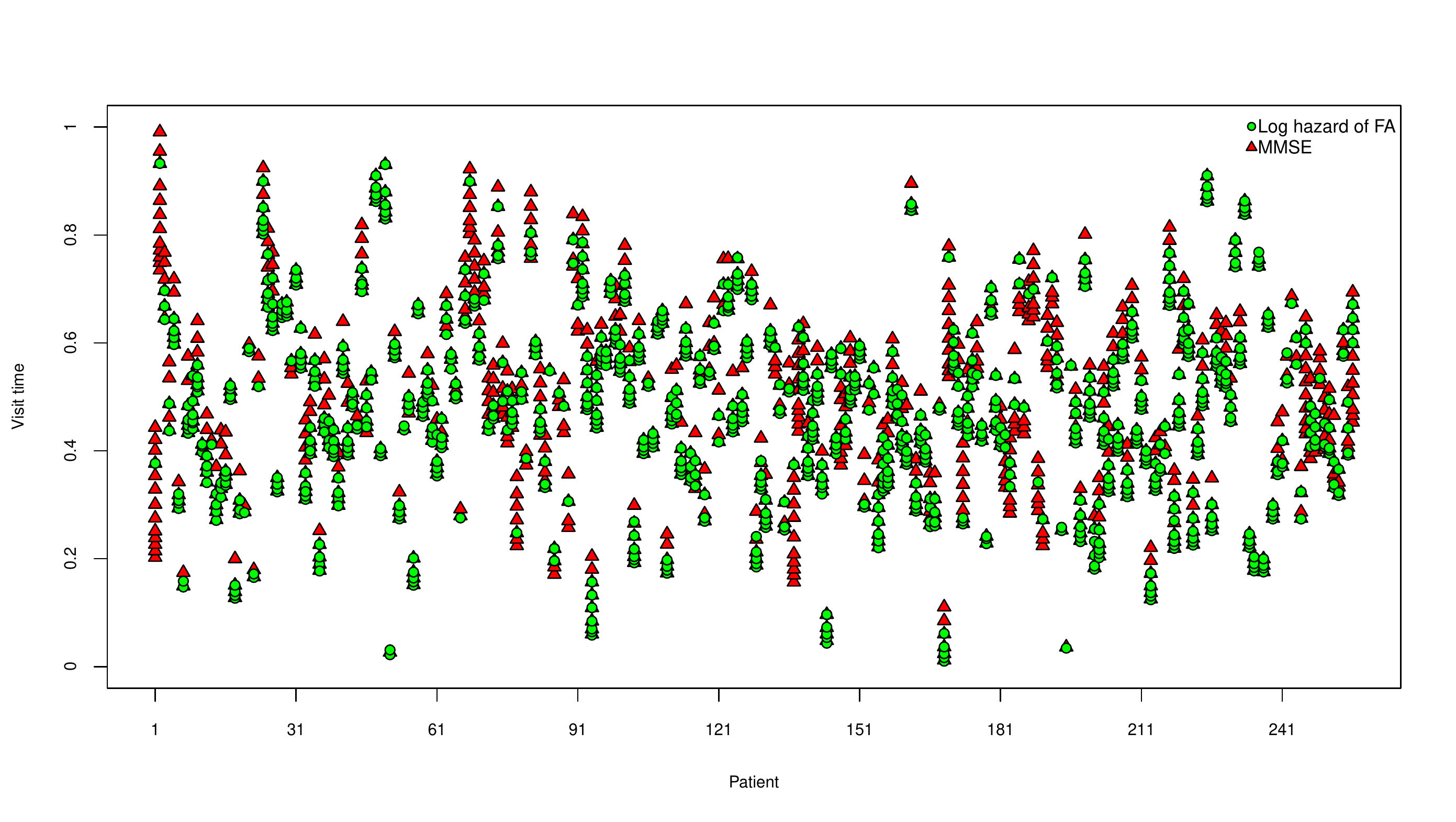}
		}
		\caption{\label{fig0}Asynchronous observation times of MMSE and FA.}
	\end{figure}

	Asynchronous longitudinal data arise when the longitudinal covariates and response are misaligned within individuals. An example of asynchronous longitudinal data is in the analysis of electronic health records (EHRs). EHR data consist of longitudinal medical records of a large number of patients in one or more electronic health care systems, such as vital signs, laboratory tests, procedure codes, and medications, among others \citep{lou2021}. Due to the retrospective nature of EHRs, measurement times are collected at each clinical encounter, which can be irregular, sparse, and heterogeneous across patients and asynchronous within patients. Another example comes from a dataset in the Alzheimer's Disease Neuroimaging Initiative (ADNI) study \citep{li2020}. In this dataset, $256$ subjects are followed up for $5$ years, where cognitive decline metrics, such as the Mini-Mental State Examination (MMSE) score, are mismatched with medical imaging measurement, such as fractional anisotropy (FA), which reflects fiber density, axonal diameter, and myelination in white matter, within individuals. Other covariates, such as age, education level, and the number of APOE4 genes, a genetic risk factor for Alzheimer's disease, are synchronous with longitudinal response MMSE, creating mixed synchronous and asynchronous longitudinal covariates. In Figure \ref{fig0}, we plot the visit times for MMSE scores and FA of all patients and two randomly selected patients.  
	
	For regression analysis of sparse asynchronous longitudinal data, several methods have been proposed in the literature. For example, \citet{xiong2010} proposed to bin the asynchronous longitudinal covariates to align with the longitudinal response. Classic longitudinal data analysis methods can be used afterwards. \citet{sent2013} explicitly addressed the asynchronous setting for generalized varying-coefficient models with one covariate but did not provide the theoretical properties of the estimators. \citet{cao2015} proposed a non-parametric kernel weighting approach for generalized linear models to address the asynchronous structure and rigorously established the consistency and asymptotic normality of the resulting estimators. This was extended to a more general set up in \citet{cao2016} and a partially linear model in \citet{chen2017}. \citet{sun2021} studied an informative measurement process for asynchronous longitudinal data. \citet{li2020} proposed a class of generalized functional partial-linear varying-coefficient models for temporally asynchronous functional neuroimaging data. These methods assume that all longitudinal covariates have the same measurement times that are asynchronous with the longitudinal response. The problem of mixed synchronous and asynchronous longitudinal covariates has not been considered before.
	
	In this paper, we propose statistical methods for estimation and inference of mixed synchronous and asynchronous longitudinal covariates in varying-coefficient models. There is a huge literature on varying-coefficient models \citep{hastie1993}. For example, \cite{fan2007} considered a semiparametric varying-coefficient partially linear model for synchronous longitudinal data. The main methodological and theoretical developments on varying-coefficient models are summarized in \cite{fan2008}. Unlike these classic results, the data structure we work with is very different. 
	The longitudinal covariates have two parts, one part is measured synchronously with the longitudinal response, and another part is asynchronous with the longitudinal response. We consider a dynamic model that allows the covariates' effect to change with respect to time. To evaluate the overall pattern and magnitude of the time-dependent coefficient or to test whether certain parametric functions are adequate in describing the overall trend of the regression relationship over time, it is more informative to construct simultaneous confidence bands (SCBs) of the non-parametric regression function. Theoretically, to construct SCBs, one has to establish the limiting distribution of the maximum deviation between the estimated and the true function. Due to the sparse, irregular, and asynchronous longitudinal data structure, one has to perform uniform investigations into a dependent empirical process where stochastic variations have to be controlled in time, synchronous longitudinal process, and asynchronous longitudinal process. Progress made in SCB for longitudinal data can be found in \citet{ma2012}, \citet{gu2014}, \citet{zheng2014}, and \citet{cao2018}, among others. For time series data, the simultaneous inference problem is studied in \citet{zhou2010} and \citet{liu2010}, among others. Due to the slow rate of convergence of the extreme value distribution, a resampling method, such as wild bootstrap, is generally adopted in SCB construction \citep{zhu2007,zhu2012}. The theoretical guarantees of wild bootstrap can be found in \citet{wu1986}, \citet{liu1988} and Chapter 2.9 of \cite{vw1996}.
	
	To improve efficiency of the varying-coefficient estimation of synchronous longitudinal covariates, we propose a two-step method. In the first step, we regress the response to the synchronous longitudinal covariates; in the second step, the residuals from the first step are regressed to the asynchronous longitudinal covariates. By avoiding unnecessary smoothing of the synchronous longitudinal covariates, we get the non-parametric rate of convergence $n^{2/5}$ for their time-varying coefficient estimation, which is faster than $n^{1/3}$ rate of convergence with a method that estimates mixed synchronous and asynchronous longitudinal covariates simultaneously. The trade-off is that the two-step method requires that the synchronous and asynchronous longitudinal covariates be uncorrelated at all time whereas the one-step method does not impose such an assumption.  
	
	The remainder of the paper is organized as follows. Section 2 establishes methods and asymptotic theory for varying-coefficient estimation of mixed synchronous and asynchronous longitudinal covariates using one-step and two-step methods, respectively. We also develop an inferential strategy with simultaneous confidence band for uncertainty quantification of the non-parametric regression function. Section 3 studies the finite-sample performance of the proposed method through simulations. Section 4 analyzes a data set from the ADNI study. Section 5 gives some concluding remarks. All technical proofs are relegated in the Appendix. 
	
	\section{Estimation and inference}\label{2.sec}
	Suppose we have a longitudinal study with $n$ subjects. For the $i$th subject, denote $Y_i(t)$ as the response variable at time $t,$ let $X_i(t)$ be a $p \times 1 $ vector of time-dependent covariate and $Z_i(t)$ be a $q \times 1 $ vector of time-dependent covariate, respectively. Mixed synchronous and asynchronous longitudinal covariates for the $i$th subject refer to the observation of $\{ Y_i(T_{ij}), X_i(T_{ij}), Z_i(S_{ik}) \}$, $j = 1, \ldots, L_i; k = 1, \ldots, M_i,$ where $T_{ij}, j =1, \ldots, L_i,$ are the observation times for the longitudinal response and synchronous longitudinal covariates and $S_{ik}, k =1, \ldots, M_i,$ are the observation times for the asynchronous longitudinal covariates with the requirement that $L_i$ and $M_i$ are finite with probability $1.$ We use univariate and bivariate counting processes to represent the observation times. Specifically, for subject $i=1, \ldots, n,$ $N_i(t) = \sum\limits_{j=1}^{L_i} I(T_{ij} \leq t)$ counts the number of observation times up to $t$ in the response and synchronous longitudinal covariates and $N^*_i(t, s) = \sum\limits_{j=1}^{L_i} \sum\limits_{k=1}^{M_i} I(T_{ij} \leq t, S_{ik} \leq s)$ counts the number of observation times up to $t$ on the response and synchronous longitudinal covariates and up to $s$ in the asynchronous longitudinal covariates \citep{cao2015,cao2016}. 
	
	We consider the following varying-coefficient model:
	\begin{equation} \label{model1}
		E \left\{Y(t) \mid X(t), Z(t)\right\} = X(t)^T \beta(t) + Z(t)^T \gamma(t),
	\end{equation}
	where $\beta(t)$ and $\gamma(t)$ are non-parametric regression functions for synchronous and asynchronous longitudinal covariates, respectively. We are interested in the statistical estimation and inference of them. 
	
	\subsection{A one-step approach}\label{2.1sec}
	To estimate $\beta(t)$ and $\gamma(t)$ simultaneously, we consider a local linear estimator \citep{fan1996,cao2018}: 
	\begin{equation}
		\begin{split}
			& \{\hat{\beta}_w(t), \hat{\dot{\beta}}_w(t), \hat{\gamma}_w(t), \hat{\dot{\gamma}}_w(t)\} \\
			= & \argmin_{\{\beta_0, \beta_1, \gamma_0, \gamma_1\}} \sum_{i=1}^{n} \iint K_{h_1, h_2} (t_1 - t, t_2 - t) \left\{ Y_i(t_1) - X_i(t_1)^T \beta_0 - X_i(t_1)^T \beta_1(t_1 - t) \right.\\
			& \left. \ \ \quad \quad \quad \quad \quad \quad \quad - Z_i(t_2)^T \gamma_0 - Z_i(t_2)^T  \gamma_1(t_2 - t) \right\}^2 dN^*_i(t_1, t_2).		\end{split}
		\label{equation1}
	\end{equation}
	
	For any fixed time point $t,$ denote $R(t_1, t_2, t) = \{ X(t_1)^T, X(t_1)^T (t_1 - t), Z(t_2)^T, Z(t_2)^T (t_2 - t) \}^T$, and $\rho_0(t) \triangleq \{ \beta_0(t)^T, \dot{\beta}_0(t)^T, \gamma_0(t)^T, \dot{\gamma}_0(t)^T \}^T,$ where $\beta_0(t)$ and $\gamma_0(t)$ are the true non-parametric regression functions and $\dot{\beta}_0(t)$ and $\dot{\gamma}_0(t)$ are the corresponding derivative functions. Assume that the conditional variance function $\mbox{var} \left\{ Y(t) \mid X(t), Z(t) \right\} = \sigma \left\{ t, X(t), Z(t) \right\}^2 < \infty.$ We need the following assumptions. 
	\begin{enumerate} 
		\item[(A1)] $N^*_i(t, s)$ is independent of $\{X_i(t), Z_i(t), Y_i(t)\}$ and $E\left\{ dN^*_i(t, s) \right\} = \eta(t, s)dtds,$ $i = 1, \ldots, n,$ where $\eta(t, s)$ is a twice continuously differentiable function for any $(t, s) \in [0, 1]^{\otimes2}.$ For $t_1 \ne s_1, t_2 \ne s_2, { P}(dN^*(t_1, t_2) = 1 \mid N^*(s_1, s_2) - N^*(s_1-, s_2-) = 1) = f(t_1, t_2, s_1, s_2)$ $dt_1dt_2$ where $f(t_1, t_2, s_1, s_2)$ is continuous for $t_1 \neq s_1, t_2 \neq s_2$ and \newline $f (t_1\pm, t_2\pm, s_1\pm, s_2\pm)$ exist.
		\item[(A2)] $\rho_0(t)$ is twice continuously differentiable for $\forall t \in [0, 1].$ For any fixed time point $t,$ $E \left\{ R(t_1, t_2, t) R(s_1, s_2, t)^T \right\} \in {\mathbb R}^{2(p+q)\times 2(p+q)}$ and ${ E}\left\{ R(t_1, t_2, t) \right\} \in {\mathbb R}^{2(p+q)}$ are twice continuously differentiable for $(t_1, t_2, s_1, s_2) \in [0, 1]^{\otimes4}.$ $E \left\{ C(t, t) C(t, t)^T \right\} \eta(t, t) \in {\mathbb R}^{(p+q) \times (p+q)}$ is positive definite for $\forall t \in [0,1],$ where $C(t, t) = \left\{ X(t)^T, Z(t)^T \right\}^T$. Moreover,
		$$
		\|E \left[ C(t, t) C(t, t)^T \sigma \left\{ t, X(t), Z(t) \right\}^2\right]\|_{\infty} \eta(t, t)< \infty, \quad \forall t \in [0,1],
		$$
		where for a square matrix $A,$ $\| A \|_{\infty} = \max_{1 \le i \le n}\sum_{j=1}^{n}|a_{ij}|. $ 
		\item[(A3)] $K(\cdot, \cdot)$ is a bivariate density function satisfying $\iint z_1 K(z_1, z_2) dz_1dz_2 = \iint z_2 K(z_1, z_2)$ \newline$dz_1dz_2 = 0, \iint K(z_1, z_2)^2 dz_1dz_2 < \infty, \iint z_1^2 K(z_1, z_2) dz_1dz_2 < \infty, \iint z_2^2 K(z_1, z_2) dz_1dz_2 < \infty$ and $\iint z_1^2 z_2^2 K(z_1, z_2) dz_1dz_2 < \infty.$
		\item[(A4)] $(nh_1h_2)^{1/2} (h_1^2 + h_2^2) \to 0$ and $nh_1h_2 \to \infty.$
	\end{enumerate}
	
	Condition (A1) requires that the observation process be independent of the covariate processes. Analogous assumptions have been made in \citet{lin2001} and \citet{cao2015}. Condition (A2) ensures the identifiability of $\rho_0 (t)$ at any fixed time point $t,$ and some smoothness assumptions on the expectation of certain functional of $R (t_1, t_2, t)$ and $\rho_0 (t).$ Conditions (A3) and (A4) specify valid kernels and bandwidths. For the bivariate kernel function, we recommend kernel functions with bounded support, such as the product of the univariate Epanechnikov kernel.
	
	The following theorem establishes the asymptotic results of $\hat{\beta}_w(t)$ and $\hat{\gamma}_w(t).$ The proofs are relegated in the Appendix. 
	\begin{theorem}\label{thm1}
		Under conditions (A1)-(A4), as $n \rightarrow \infty,$ we have 

		$$\sqrt{nh_1h_2} \left[ \{ \hat{\beta}_w(t) - \beta_0(t) \}^T, \{ \hat{\gamma}_w(t) - \gamma_0(t) \}^T \right]^T \stackrel{ d}{\to} N \{ 0, A^{*}(t)^{-1} \Sigma^{*}(t) A^{*}(t)^{-1} \},$$
		where $$A^{*}(t) = E \{ C(t, t) C(t,t)^T \} \eta(t, t)$$ and  
  $$ \Sigma^{*}(t) = \big\{\iint K(z_1, z_2)^2 dz_1 dz_2 \big\}E \left[  C(t, t) C(t, t)^T \sigma \{ t, X(t), Z(t) \}^2 \right]  \eta(t, t). $$ 
	\end{theorem}
	Theorem \ref{thm1} derives the joint asymptotic distribution of $\hat{\beta}_w(t)$ and $\hat{\gamma}_w(t).$ If we let $h_{\alpha}=h_1=h_2$, the bias of $\hat{\beta}_w(t)$ and $\hat{\gamma}_w(t)$ in Theorem \ref{thm1} is $O(h_{\alpha}^2)$, which is the same as that in Theorem 2 of \cite{cao2015}. The reason why the $O(h_\alpha)$ term vanishes is that we use symmetric density functions as the kernel functions, which are specified in (A3). When the sample size $n$ goes to infinity, under (A4), the bias terms vanish. In general, the off-diagonal part of the limiting variance covariance matrix is non-zero. In the special case that $E\{X(t) Z(t)^T \}=0_{p\times q}, $ $\hat{\beta}_w(t)$ and $\hat{\gamma}_w(t)$ are independent. Moreover, if we assume that $\sigma\{t, X(t), Z(t) \}^2 = \sigma(t)^2,$ then $\mbox{Var}\{\hat{\beta}_w(t)\}= [E \{X(t) X(t)^T \} \eta(t,t)]^{-1} \big\{ \iint K(z_1, z_2)^2 dz_1 dz_2\big\} \sigma(t)^2$ and $\mbox{Var}\{\hat{\gamma}_w(t) \} = [E \{Z(t) Z(t)^T \} \eta(t,t)]^{-1}\big\{\iint K(z_1, z_2)^2 dz_1 dz_2\big\}\sigma(t)^2,$ respectively.
	
	Our method depends on the proper choice of bandwidths. If we let $h_{\alpha} = h_1 = h_2,$ according to condition (A4), a valid bandwidth is larger than $O(n^{-1/2}),$ and the bias of $\hat{\beta}_w (t)$ and $\hat{\gamma}_w (t)$ in Theorem \ref{thm1} is of order $h_\alpha^2,$ so we should choose bandwidth $h_{\alpha} = o(n^{-1/6}).$ With this choice of $h_\alpha$, we obtain a rate of convergence $O(n^{1/3}),$ which is the same as that achieved in \citet{cao2015}. This is slower than the non-parametric rate of convergence $n^{2/5}$ due to the asynchronous longitudinal data structure.
	
	In Theorem \ref{thm1}, $A^*(t)$ and $\Sigma^*(t)$ are not directly computable. In practice, we use the estimating equation derived from (\ref{equation1}) and estimate the variance covariance matrix of $\hat{\rho}_w(t)$ with the sandwich formula:
	\begin{equation*}
		\begin{split}
			& \widehat{\mathrm{var}} \{ \hat{\rho}_w(t) \} \\
			=
			& \left\{ \sum_{i=1}^{n} \iint K_{h_1, h_2}(t_1 - t, t_2 - t) R_i(t_1, t_2, t) R_i(t_1, t_2, t)^T dN^*_i(t_1, t_2) \right\}^{-1} \\
			& \times \sum_{i=1}^{n} \left[ \iint K_{h_1, h_2}(t_1 - t, t_2 - t) R_i(t_1, t_2, t) \left\{ Y_i(t_1) - R_i(t_1, t_2, t)^T \hat{\rho}_w(t) \right\} dN^*_i(t_1, t_2) \right]^{\otimes 2} \\
			& \times \left\{ \sum_{i=1}^{n} \iint K_{h_1, h_2}(t_1 - t, t_2 - t) R_i(t_1, t_2, t) R_i(t_1, t_2, t)^T dN^*_i(t_1, t_2) \right\}^{-1}.
		\end{split}
	\end{equation*}
	The variance covariance matrices of $\hat{\beta}_w(t)$ and $\hat{\gamma}_w(t)$ are the upper $p \times p$ submatrix of the upper left $(2p) \times (2p)$ submatrix and the upper $q \times q$ submatrix of the lower right $(2q) \times (2q)$ submatrix of $\widehat{\mathrm{var}} \{ \hat{\rho}_w(t) \}$, respectively.
	
	\subsection{Simultaneous confidence bands}
	To get uncertainty quantification and evaluate the overall magnitude of variation of the non-parametric regression functions, we construct simultaneous confidence bands (SCB). Here, we use $\gamma(t)$ to illustrate. Specifically, for a pre-specified confidence level $1-\alpha$ and a given smooth function of $q$ dimensions $a(t),$ we aim to find smooth random functions $L(t)$ and $U(t)$, such that 
	$$
	\mathbb{P} \{L(t) \le a(t)^T \gamma(t) \le U(t), \forall t \in [a,b] \}\rightarrow 1-\alpha
	$$
	as number of subjects $n \rightarrow \infty$ for a pre-specified closed interval $[a,b].$ Theoretical results on SCB construction rely on extreme value theory with slow rate of convergence \citep{fan2000, cao2018}. We shall adopt a wild bootstrap approach to resample the residuals \citep{wu1986, liu1988}. The wild bootstrap is computationally efficient as it multiplies a random disturbance on the residuals without repeated estimation of the non-parametric regression function. Specific steps are as follows. 
	
	Step 1. For each subject $i, i =1, 2, \ldots, n,$ calculate the kernel-weighted residual
	$$\hat{r}_i(t_1, t_2, t) = K_{h_1, h_2} (t_1 - t, t_2 - t) \left\{ Y_i(t_1) - R_i(t_1, t_2, t)^T \hat{\rho}_w(t) \right\}.$$
	
	Step 2. Obtain an approximation of $\hat{\gamma}_w(t) - \gamma_0(t)$ through $$\hat{Q}_n(t)  = \frac{1}{n} \sum_{i=1}^n \left\{ f_n(t) \right\}^{-1} \iint Z_i(t_2) \hat{r}_i(t_1, t_2, t) dN^*_i(t_1, t_2),$$
	where $f_n (t) = \frac{1}{n} \sum_{i=1}^{n} \iint K_{h_1, h_2} (t_1 - t, t_2 - t) Z_i (t_2) Z_i (t_2)^T dN^*_i (t_1, t_2).$
	
	Step 3. Independently generate $\left\{ u_i, i = 1, \dots, n \right\}$ from a distribution with mean $0$ and standard deviation $1.$. For example, we can use $N(0,1)$ or the Rademacher distribution which takes values $+1$ and $-1$ with equal probability. Construct the stochastic process
	$$\hat{Q}_n^{u}(t)  = \frac{1}{n} \sum_{i=1}^n u_i \cdot \left\{ f_n(t) \right\}^{-1} \iint Z_i(t_2) \hat{r}_i(t_1, t_2, t) dN^*_i(t_1, t_2).$$
	
	Step 4. Repeat Step 3 $B$ times to obtain $\{ \sup\limits_t \mid a(t)^T\hat{Q}_n^{u(1)}(t) \mid, \dots, \sup\limits_t \mid a(t)^T\hat{Q}_n^{u(B)}(t) \mid \},$ and denote its $1 - \alpha$ empirical percentile as $c_{\alpha}.$ The SCB for $a(t)^T \gamma(t)$ can be written as $$\left( a(t)^T \hat{\gamma}_w(t) - c_{\alpha}, a(t)^T\hat{\gamma}_w(t) + c_{\alpha} \right).$$
	We recommend the Rademacher distribution as the multiplier in Step 3 through extensive simulation studies with different random variables. The details are omitted.
	
	We remark that for a given $p$-dimensional smooth function $d(t),$ the construction of SCB for $d(t)^T \beta(t)$ is exactly the same except in Step 2, let $\hat{J}_n(t) = \frac{1}{n} \sum_{i=1}^n \left\{ g_n(t) \right\}^{-1} \iint X_i(t_1) \hat{r}_i(t_1, t_2, t) \newline dN^*_i(t_1, t_2),$ where $g_n(t) = \frac{1}{n} \sum_{i=1}^{n} \iint K_{h_1, h_2} (t_1 - t, t_2 - t) X_i (t_1) X_i (t_1)^T dN^*_i (t_1, t_2).$ The remaining steps are based on $\hat{J}_n(t)$ and we replace $a(t)$ by $d(t).$
	
	\subsection{A two-step approach}\label{2.2sec}
	In (\ref{model1}), $\beta (t)$ is the regression function of the synchronous longitudinal covariate. In classic non-parametric regression analysis of longitudinal data, we get the $O(n^{2/5})$ rate of convergence. Can we achieve the same rate of convergence  as the estimation of $\beta(t)$ with mixed synchronous and asynchronous longitudinal covariates? In this subsection, we provide an affirmative answer to this question.
	
	Specifically, we propose a two-step approach to estimate $\beta(t)$ and $\gamma(t),$ respectively. In the first step, we aim to obtain unbiased estimation of $\beta(t)$ from (\ref{model1}) using only $X(t)$ and $Y(t)$. There are two strategies to handle the part that involves $\gamma(t).$ We can either remove it or take it into account by a non-parametric function. In the second step, we regress residuals from the first step on the asynchronous longitudinal covariates $Z(t)$ to get regression function estimation of $\gamma(t)$ similar to that in \cite{cao2015}. 
	
	We need a critical assumption \begin{equation}\label{orthogonality}
	   E\{Z(t) \mid X(t) \} = E \{ Z(t)\}.
	\end{equation}
Condition (\ref{orthogonality}) states that $X(t)$ and $Z(t)$ are uncorrelated when $X(t) \ne 0$ for any $t.$ This condition is plausible when $X(t)$ indicates a time-invariant treatment, which is randomized and independent of other covariates. This condition means that there is no unmeasured confounder between treatment and outcome. This assumption can be strong in observational studies, where unmeasured confounders abound. In classic linear models, when the covariates are orthogonal with each other, the estimation of regression coefficients remains unbiased with a larger variance if some covariates are omitted. Unlike linear models, bias occurs when orthogonal covariates are omitted for longitudinal data if a non-parametric intercept is not added.  
If $X(t)$ and $Z(t)$ are correlated, there will be bias using the proposed two-step method.

	{\it A centering approach to estimate $\beta(t)$}. \quad From (\ref{model1}), taking a conditional expectation on $X(t),$ by (\ref{orthogonality}), we have 
	\begin{equation}
		E \{ {Y}(t) \mid X(t)\} = {X}(t)^T \beta(t) + \{E Z(t)\}^T \gamma(t) \label{2022-3-28}.
	\end{equation}
	Taking another expectation, we get unconditional expectation of $Y(t)$ as
	\begin{equation} 
		E \{Y(t)\} = \{E X(t)\}^T \beta(t) + \{E Z(t)\}^T \gamma(t). \label{2022-3-28-2}
	\end{equation}
	By taking the difference between (\ref{2022-3-28}) and (\ref{2022-3-28-2}), we obtain 
	\begin{equation}
		E \{ \tilde{Y}(t) \mid \tilde{X}(t)\} = \tilde{X}(t)^T \beta(t),
		\label{model2}
	\end{equation}
	where $\tilde{Y}(t) = Y(t) - E \{ Y(t) \}$ and $\tilde{X}(t) = X(t) - E \{ X(t) \}.$ Unbiased estimation of $\beta(t)$ can be obtained through (\ref{model2}). In (\ref{model2}), $m_Y(t) = E\{Y(t)\}$ and $m_X(t) = E \{X(t) \}$ are not directly observed. We propose Nadaraya-Watson type of estimators as follows \citep{nadaraya1964, watson1964}. Let $\hat{m}_Y(t_0) = \{\sum\limits_{i=1}^{n} \int K_h(t-t_0) Y_i(t) dN_i(t)\}/\{\sum\limits_{i=1}^{n} \int K_h(t-t_0) dN_i(t)\},$ and $\hat{m}_X(t_0) = \{\sum\limits_{i=1}^{n} \int K_h(t-t_0) X_i(t) dN_i(t)\}/\{\sum\limits_{i=1}^{n} \int K_h(t-t_0) dN_i(t)\},$ where $K(\cdot)$ is a kernel function with bounded support,  $K(\cdot) \ge 0, \int_{-\infty}^\infty K(t)dt = 1$ and $K_h(t) = h^{-1} K(t/h).$ The bandwidth $h \rightarrow 0$ and $nh \rightarrow \infty.$ Denote $\hat{Y}_i(t) = Y_i(t) - \hat{m}_Y(t)$ and $\hat{X}_i(t) = X_i(t) - \hat{m}_X(t).$ We aim to find 
	\begin{equation}
		\begin{split}
			& \{\hat{\beta}_c(t), \hat{\dot{\beta}}_c(t)  \} \\
			= & \argmin_{\beta_0, \beta_1 \in R^p}	\Big[ \frac{1}{n} \sum_{i=1}^{n} \int K_h(t_1 - t) \left\{ \hat{Y}_i(t_1) - \hat{X}_i(t_1)^T \beta_0 - \hat{X}_i(t_1)^T \beta_1(t_1 - t) \right\}^2 dN_i(t_1) \Big].
			\label{5}
		\end{split}
	\end{equation}
	Define
	\begin{equation*}
		\begin{split}
			S_{n, l}(t) & = \frac{1}{nh} \sum\limits_{i=1}^n \int K_h(t_1 - t) \hat{X}_i(t_1) \hat{X}_i(t_1)^T \{(t_1 - t)/h\}^{l}dN_i(t_1), l = 0, 1, 2\\
			\mbox{and} \quad q_{n, l}(t) & = \frac{1}{nh} \sum\limits_{i=1}^n \int K_h(t_1 - t) \hat{X}_i(t_1) \hat{Y}_i(t_1) \{ (t_1 - t)/h\}^{l}dN_i(t_1), l = 0, 1.
		\end{split}
	\end{equation*}
	Then
	$$\begin{pmatrix}
		\hat{\beta}_c (t)\\
		h\hat{\dot{\beta}}_c (t)
	\end{pmatrix} = \begin{pmatrix}
		S_{n,0} & S_{n,1}\\
		S_{n,1} & S_{n,2}
	\end{pmatrix}^{-1} \begin{pmatrix}
		q_{n,0}\\
		q_{n,1}
	\end{pmatrix}.$$
	Next, we show the asymptotic properties of $\hat{\beta}_c(t)$. Denote $\tilde{W}_i(t_1, t) = \{\tilde{X}_i(t_1)^T, \tilde{X}_i(t_1)^T (t_1 - t)\}^T,$ $\theta_0(t) = \{ \beta_0(t)^T, \dot{\beta}_0(t)^T \}^T,$ $\hat{\theta}_c(t) = \{ \hat{\beta}_c(t)^T, \hat{\dot{\beta}}_c(t)^T \}^T,$ and $\mbox{var} \{\tilde{Y}(t)\mid\tilde{X}(t)\} = \sigma \{t, \tilde{X}(t)\}^2.$ We need the following assumptions. 
	
	\begin{enumerate}
		\item [(A5)] $N_i(t)$ is independent of $\left\{ X_i(t), Y_i(t) \right\},$ and $E \{ dN_i(t) \} = \lambda(t) dt, i = 1, \dots, n,$ where $\lambda(t)$ is twice continuously differentiable for $\forall t \in [0, 1].$
		
		\item [(A6)] $\theta_0(t)$ is twice continuously differentiable for $\forall t \in [0, 1]$. For any fixed time point $t,$ $E\{ \tilde{W}(t_1, t) \}$ and $E\{ \tilde{W}(t_1, t) \tilde{W}(s_1, t)^T \}$ are twice continuously differentiable for $(t_1, s_1) \in [0, 1]^{\otimes2}$. 
		Furthermore, $E \{ \tilde{X}(t) \tilde{X}(t)^T \} \lambda(t)$ is positive definite for $\forall t \in [0, 1]$. Furthermore,
		$$\|E [ \tilde{X}(t) \tilde{X}(t)^T \sigma \{t, \tilde{X}(t)\}^2 ]\|_{\infty} \lambda(t) < \infty,$$
		where for a square matrix $A,$ $\|A\|_{\infty} = \max_{1\le i \le n}\sum_{j=1}^{n}|a_{ij}|.$
		\item [(A7)] $K(\cdot)$ is a symmetric density function satisfying $\int z K(z) dz = 0,$ $\int z^2 K(z) dz < \infty$ and $\int K(z)^2 dz < \infty.$
		
		\item[(A8)] $nh \to \infty$ and $nh^5 \to 0.$
	\end{enumerate}
	
	Conditions (A5)-(A8) are similar in spirit to conditions (A1)-(A4) in Section \ref{2.1sec}. (A5) is the univariate version of (A1). (A6) is a modified identifiability condition for $\theta_0(t).$ (A7) and (A8) are requirements on the kernel function and the bandwidth. 
	
	We state the asymptotic distribution of $\hat{\beta}_c(t)$ in the following theorem and relegate the proofs in the Appendix.
	
	\begin{theorem} \label{thm2}
		Under conditions (\ref{orthogonality}), (A5)-(A8), as $n\rightarrow \infty,$ we have  
		$$\sqrt{nh} \{\hat{\beta}_c(t) - \beta_0(t)\} \stackrel{ d}{\to} N \{0, A^{-1}(t) \Sigma(t) A^{-1}(t)\},$$
		where
		\begin{equation*}
			\begin{split}
				A(t) & = E\{\tilde{X}(t) \tilde{X}(t)^T\} \lambda(t)\\
				\mbox{and} \quad \Sigma(t) & = \big\{\int K(z)^2 dz \big\} E \left[ \tilde{X}(t) \tilde{X}(t)^T \sigma \{t, \tilde{X}(t)\}^2 \right] \lambda(t).
			\end{split}
		\end{equation*}
	\end{theorem}
	
We need one bandwidth $h$ and the bias of $\beta_c(t)$ in Theorem \ref{thm2} is $O(h^2)$, which is the same as in classic nonparametric regression such as Theorem 1 of \cite{fan2008}. Our method depends on the selection of the bandwidth, which strikes a balance between bias and variability. From (A8), we shall choose the bandwidth $h = o(n^{-1/5}).$ With this choice of the bandwidth, we achieve a convergence rate $O(n^{2/5}),$ the same as non-parametric regression of classic longitudinal data \citep{fan2008} and faster than that obtained in Theorem \ref{thm1}. It is counterintuitive that we get an improved rate of convergence with less information -- we do not use any information contained in $Z_i(S_{ik}), i = 1, \ldots, n; k = 1, \ldots, M_i.$ The key is the orthogonality assumption imposed in (\ref{orthogonality}), which allows us to obtain an unbiased estimation of the non-parametric regression function $\beta_0(t).$ When (\ref{orthogonality}) is violated, we can regress the residuals of $Y(t)$ on $Z(t)$ on the residuals of $X(t)$ on $Z(t)$ similar to the FWL theorem in linear models \citep{frisch1933, lovell1963, ding2021}. We conjecture that the results would be similar to the simultaneous estimation of $\beta(t)$ and $\gamma(t),$ with a slower convergence rate for the estimation of $\beta(t).$ 
	
	For statistical inference, we need to estimate the variance of $\hat{\beta}_c(t).$ We use the sandwich formula to estimate the variance covariance matrix of $\{\hat{\beta}_c(t), \hat{\dot{\beta}}_c(t) \}$ as follows: 
	\begin{equation*}
		\begin{split}
			\widehat{\mathrm{var}} \{ \hat{\theta}_c(t) \} = & 
			\left\{ \sum_{i=1}^{n} \iint K_{h}(t_1 - t) \hat{W}_i(t_1, t) \hat{W}_i(t_1, t)^T dN_i(t_1) \right\}^{-1} \\
			& \times \sum_{i=1}^{n} \left[ \iint K_{h}(t_1 - t) \hat{W}_i(t_1, t) \left\{ \hat{Y}_i(t_1) - \hat{W}_i(t_1, t)^T \hat{\theta}_c(t) \right\} dN_i(t_1) \right]^{\otimes 2} \\
			& \times \left\{ \sum_{i=1}^{n} \iint K_{h}(t_1 - t) \hat{W}_i(t_1, t) \hat{W}_i(t_1, t)^T dN_i(t_1) \right\}^{-1},
		\end{split}
	\end{equation*}
	where $\hat{W}_i (t_1, t) = \{ \hat{X}_i(t_1)^T, \hat{X}_i(t_1)^T (t_1 - t) \}^T, \hat{X}_i(t) = X_i(t) - \hat{m}_X(t),$ and $\hat{Y}_i(t) = Y_i(t) - \hat{m}_Y(t).$ The variance covariance matrix of $\hat{\beta}_c(t)$ is the upper left $p\times p$ submatrix of $\widehat{\mathrm{var}} \{ \hat{\theta}_c(t) \}$ while we treat $\dot{\beta}_c(t)$ as a nuisance function. In the literature on density estimation, $\hat{\dot{\beta}}_c(t)$ estimated from (\ref{5}) has more desirable properties than conventional kernel density estimation. More details on this can be found in \citet{cattaneo2020}. 
	
	{\it A varying-coefficient model approach to estimate $\beta(t)$}. \quad Another strategy to obtain an unbiased estimation of $\beta(t)$ is to absorb the part involving $\gamma(t)$ through a non-parametric function. We use (\ref{2022-3-28-2}) and treat $E\{Z(t) \}^T \gamma(t)$ as a non-parametric intercept. Specifically, consider
	\begin{equation}
		E \{Y(t) \mid X(t)\} = X^{*}(t)^{T} \beta^{*}(t),
		\label{model3}
	\end{equation}
	where $X^{*}(t) = \{ 1, X(t)^{T} \}^{T},$ and $\beta^{*}(t) = \{ \alpha(t), \beta(t)^T \}^{T},$ where $\alpha(t)$ and $\beta(t)$ are non-parametric regression functions. Our aim is to find
	\begin{equation}
		\begin{split}
			&	\{ \hat{\beta}^{*}_v (t), \hat{\dot{\beta}}^{*}_v (t) \} \\
			= & \argmin_{\beta_0^{*}, \beta_1^{*} \in R^{p+1}} \sum_{i=1}^{n} \int K_{h} (t_1 - t) \left\{ Y_i (t_1) - X_i^{*} (t_1)^T \beta_0^{*} - X_i^{*} (t_1)^T \beta_1^{*} (t_1 - t) \right\}^2 dN_i (t_1),
		\end{split}
		\label{7}
	\end{equation}
	where $K \left( \cdot \right)$ is a kernel function, $K_{h} (t) = K \left( t/h \right)/h,$ and $h$ is a bandwidth.
	
	Next we present the asymptotic distribution of $\hat{\beta}_v(t)$ and relegate the proofs in the Appendix. 
	\begin{theorem} \label{thm3}
		Under conditions (\ref{orthogonality}), (A5)-(A8), as $n \rightarrow \infty,$ we have 
		$$
		\sqrt{nh} \{ \hat{\beta}_v(t) - \beta_0(t) \} \stackrel{d}{\to} N \{ 0, A^{-1}(t) \Sigma(t) A^{-1}(t)\},
		$$
		where $A(t)$ and $\Sigma(t)$ are specified in Theorem \ref{thm2}.
	\end{theorem}
The asymptotic distribution and the bias of $\hat{\beta}_v(t)$ is the same as that of $\hat{\beta}_c(t).$ Simulation studies provide numerical support for this. For variance estimation, let $Q_i(t_1, t) = \{1, X_i(t_1)^{T}, (t_1 - t), X_i(t_1)^T (t_1 -t)\}^T$ and $\hat{\iota}_v(t) \triangleq \{ \hat{\alpha}_v(t), \hat{\beta}_v(t)^T, \hat{\dot{\alpha}}_v(t), \hat{\dot{\beta}}_v(t)^T \}^T.$ We calculate the variance of $\hat{\iota}_v(t)$ by the following formula:
	\begin{equation*}
		\begin{split}
			\widehat{\mathrm{var}} \{ \hat{\iota}_v(t) \} = & \left\{ \sum_{i=1}^{n} \iint K_{h}(t_1 - t) \hat{Q}_i(t_1, t) \hat{Q}_i(t_1, t)^T dN_i(t_1) \right\}^{-1} \\
			& \times \sum_{i=1}^{n} \left[ \iint K_{h}(t_1 - t) \hat{Q}_i(t_1, t) \left\{ \hat{Y}_i(t_1) - \hat{Q}_i(t_1, t)^T \hat{\iota}_v(t) \right\} dN_i(t_1) \right]^{\otimes 2} \\
			& \times \left\{ \sum_{i=1}^{n} \iint K_{h}(t_1 - t) \hat{Q}_i(t_1, t) \hat{Q}_i(t_1, t)^T dN_i(t_1) \right\}^{-1}.
		\end{split}
	\end{equation*}
	The variance covariance matrix of $\{\hat{\alpha}_v(t), \hat{\beta}_v(t)^T \}^T$ is the upper $(p+1) \times (p+1)$ submatrix of $\widehat{\mathrm{var}} \{ \hat{\iota}_v(t) \}.$ There is a trade-off between the centering approach and the non-parametric intercept approach. In terms of computation, the centering approach is faster. If we are interested in the non-parametric intercept term, the varying-coefficient model is more informative. 
	
	{\it A kernel weighting approach to estimate $\gamma(t)$}. \quad In the first stage, $\beta(t)$ can be estimated using either the centering or the varying-coefficient model approach, and they have the same asymptotic distribution. Once $\hat{\beta}(t)$ is obtained, we can estimate $\gamma(t)$ by regressing the residuals from the first step on the asynchronous longitudinal covariates $Z(t).$ Due to the asynchronicity of $Z(t),$ we need two bandwidths in the calculation of $\hat{\gamma}(t).$ Specifically, 
	\begin{equation}
		\begin{split}
			\{ \hat{\gamma}(t), \hat{\dot{\gamma}}(t) \} = \argmin_{\gamma_0, \gamma_1 \in R^q} \sum_{i=1}^{n} \iint & K_{h_1, h_2}(t_1 - t, t_2 - t) \{ Y_i(t_1) - X_i(t_1)^T \hat{\beta}(t_1) \\
			& - Z_i(t_2)^T \gamma_0 - Z_i(t_2)^T \gamma_1(t_2 - t) \}^2 dN^*_i(t_1, t_2),
		\end{split}
		\label{9}
	\end{equation}
	where $K_{h_1, h_2}(t,s) = K(t/h_1, s/h_2)/(h_1h_2),$ where $K(t,s)$ is a bivariate kernel function, which is usually taken to be the product of the univariate kernel function and $h_1$ and $h_2$ are bandwidths. As $\hat{\beta}(t)$ has a faster convergence rate, whether we use the true $\beta(t)$ or $\hat{\beta}(t)$ does not affect the inference of $\hat{\gamma}(t).$
	
	We establish the asymptotic distribution of $\hat{\gamma}(t)$ in the following theorem and relegate the proofs to the Appendix.
	
	\begin{theorem} \label{thm4}
		Under conditions (\ref{orthogonality}), (A1)-(A4), as $n\rightarrow \infty,$ we have 
$$
		\sqrt{nh_1h_2} \{ \hat{\gamma}(t) - \gamma_0(t) \} \stackrel{d}{\to} N \{ 0, A^{+}(t)^{-1} \Sigma^{+}(t) A^{+}(t)^{-1} \},
		$$
		where 
		\begin{equation*}
			\begin{split}
				A^{+}(t) & = E \{ Z(t) Z(t)^T \} \eta(t, t)\\
				\mbox{and} \quad \Sigma^{+}(t) & = \big\{\iint K(z_1, z_2)^2 dz_1 dz_2\big\} E\left[ Z(t) Z(t)^T \sigma \{ t, X(t), Z(t) \}^2\right] \eta(t, t).
			\end{split}
		\end{equation*}
	\end{theorem}

In the second step, we need two bandwidths. If we set $h_1=h_2=h_{\beta},$ the bias of $\hat{\gamma}(t)$ in Theorem \ref{thm4} is $O(h_\beta^2).$ We achieve the same rate of convergence $O(n^{1/3})$ for the estimation of $\hat{\gamma}(t)$ as in the one-step method. This is the price we pay for the asynchronous longitudinal data structure. Whether we can improve this rate of convergence remains open at the moment. 

	Let $\hat{\phi} (t) = \{ \hat{\gamma}(t)^T, \hat{\dot{\gamma}}(t)^T \}^T,$ $V_i (t_2, t) = \{ Z_i (t_2)^T, Z_i (t_2)^T (t_2 - t) \}^T,$ we can calculate 
$$
	\begin{aligned}
		& \widehat{\mathrm{var}} \{ \hat{\phi}(t) \} \\
		& = \left\{ \sum_{i=1}^{n} \iint K_{h_1, h_2}(t_1 - t, t_2 - t) V_i(t_2, t) V_i(t_2, t)^T dN^*_i(t_1, t_2) \right\}^{-1} \\
		& \times \sum_{i=1}^{n} \left[ \iint K_{h_1, h_2}(t_1 - t, t_2 - t) V_i(t_2, t) \left\{ Y_i(t_1) - X_i(t_1)^T \hat{\beta}(t_1) - V_i(t_2, t)^T \hat{\phi}(t) \right\} dN^*_i(t_1, t_2) \right]^{\otimes 2} \\
		& \times \left\{ \sum_{i=1}^{n} \iint K_{h_1, h_2}(t_1 - t, t_2 - t) V_i(t_2, t) V_i(t_2, t)^T dN^*_i(t_1, t_2) \right\}^{-1}.
	\end{aligned}
 $$
	The variance covariance matrix of $\hat{\gamma}(t)$ is the upper $p \times p$ submatrix of $\widehat{\mathrm{var}} \{ \hat{\phi}(t) \}.$ Statistical inference can be carried out afterwards. 
	
	\subsection{Bandwidth selection}\label{2.3sec}
	Our methods depend critically on the choice of the bandwidth. Asymptotic results provide theoretical range of valid bandwidths. In practice, it is desirable to perform a data-driven bandwidth selection. \citet{cao2015} and \citet{chen2017} used a data-splitting strategy to separately calculate bias and variability and choose a bandwidth that minimizes the mean squared error.
	
	In this paper, we use a kernel smoothed $D$ folds cross-validation procedure to select the optimal bandwidth. We use the two-step approach with centering in the first step and kernel weighting in the second step to illustrate this point. Specifically, we split the data set into $D$ folds, and estimate the non-parametric regression functions withholding one fold. The prediction error is calculated based on the one-fold data withheld and the regression functions are estimated with the other $D-1$ fold data. We do this $D$ times and the optimal bandwidth is the one that minimizes the average squared prediction error for any fixed time point $t.$
	
	Specifically, for $\hat{\beta}(t)$ at the time point $t,$ the average squared prediction error is 
	\begin{equation}
		\mathrm{ASPE}_{\hat{\beta}(t)} (h; t) = \frac{1}{ D } \sum\limits_{k=1}^{D } A_{n^{(k)}}(h; t)/B_{n^{(k)}}(h; t),
		\label{10}
	\end{equation}
	where
	\begin{equation*}
		\begin{split}
			A_{n^{(k)}}(h; t) & = \sum\limits_{i=1}^{n^{(k)}} \int K_h(t_1 - t) \{ \hat{Y}_i(t_1) - \hat{X}_i(t_1)^T \hat{\beta}^{(-k)}(t_1) \}^2 dN_i(t_1),\\
			\mbox{and} \quad 
			B_{n^{(k)}}(h; t) & = \sum\limits_{i=1}^{n^{(k)}} \int K_h(t_1 - t) dN_i(t_1), \quad k=1,\ldots, D,
		\end{split}
	\end{equation*}	
	where $n^{(k)}$ is the number of subjects in the $k$th fold, $\hat{\beta}^{(-k)}(t)$ is estimated without the subjects in the $k$th fold, $\hat{Y}_i(t)$ estimates $\tilde{Y}_i(t)$, $\hat{X}_i(t)$ estimates $\tilde{X}_i(t),$ $K_h(\cdot) = K(\cdot)/h$, $K(\cdot)$ is a kernel density function and $h$ is the bandwidth. Similarly, for $\hat{\gamma}(t)$ at time point $t,$ the average squared prediction error is 
	\begin{equation}
		\mathrm{ASPE}_{\hat{\gamma}(t)} (h_1, h_2; t) = \frac{1}{D} \sum\limits_{k=1}^{D} C_{n^{(k)}}(h_1, h_2; t)/D_{n^{(k)}}(h_1, h_2; t),
		\label{11}
	\end{equation}
	where
	\begin{equation*}
		\begin{split}
			C_{n^{(k)}}(h_1, h_2; t) = & \sum\limits_{i=1}^{n^{(k)}} \iint K_{h_1, h_2}(t_1 - t, t_2 - t) \\
			& \times \left\{ Y_i(t_1) - X_i(t_1)^T \hat{\beta}(t_1) - Z_i(t_2)^T \hat{\gamma}^{(-k)}(t_2) \right\}^2  dN^*_i(t_1, t_2),\\
			\mbox{and} \quad		D_{n^{(k)}}(h_1, h_2; t) = & \sum\limits_{i=1}^{n^{(k)}} \iint K_{h_1, h_2}(t_1 - t, t_2 - t) dN^*_i(t_1, t_2),
		\end{split}
	\end{equation*}
	where $n^{(k)}$ is the number of subjects in the $k$th fold, $\hat{\gamma}^{(-k)}(t)$ is estimated without the subjects in the $k$th fold, $\hat{\beta}(t)$ is estimated from Step 1 with the optimal bandwidth plugged in, $K_{h_1, h_2}(\cdot, \cdot)$ is a bivariate kernel function, and $h_1$ and $h_2$ are bandwidths. It is worth noting that when using the two-step approach, we need to first select the bandwidth $h_{{opt}}$ for $\hat{\beta} (t)$ by minimizing (\ref{10}), and then select the optimal bandwidth for $\hat{\gamma} (t)$ by minimizing (\ref{11}). If we want to calculate the optimal bandwidth of the regression function in a certain range of $t,$ we can calculate the integrated ASPE.
	
	\section{Numerical studies}\label{3.sec}
	In this section, we evaluate the performance of the finite sample of the proposed method through simulations. 
	
	\subsection{Data generating process and model specification}\label{3.1sec}
	We generate $1,000$ datasets, each consisting of $400$ or $900$ subjects. For each subject, the number of observations of synchronous longitudinal data $\{ X(t), Y(t) \}$ and asynchronous longitudinal covariate $Z(t)$ follow $\mbox{Poisson}(5)+1$. The observational times are independently generated from the standard uniform distribution $\mathcal{U}(0,1)$ for $\{X(t), Y(t)\}$ and $Z(t),$ respectively. The covariate processes $X(t)$ and $Z(t)$ are both Gaussian, with $E \{X(t)\} = 0,$ $E \{Z(t)\} = 2 (t-0.5)^2$ and $\mbox{Cov} \{X(t), X(s)\} = \mbox{Cov} \{Z(t), Z(s)\} = e^{- \mid t-s \mid}.$ The values of the stochastic process $Z(t)$ are recorded at the observational times of $\{X(t), Y(t)\},$ only in the data generation stage, to generate $Y(t)$ in the model (\ref{model1}).  
	
	The longitudinal response is generated from
	$$Y(t) = X(t)^T \beta(t) + Z(t)^T \gamma(t) + \epsilon(t),$$
	where $\beta(t)$ and $\gamma(t)$ are time-dependent coefficients, and $\epsilon (t)$ is a Gaussian process, independent of $X(t)$ and $Z(t),$ with $E \{\epsilon(t)\} = 0$ and $\mbox{Cov} \{\epsilon (t), \epsilon (s)\} = 2^{- \mid t-s \mid}.$ For the time-dependent coefficients, we consider two different settings: (i) $\beta(t) = 3(t-0.4)^2, \gamma(t) = \mbox{sin}(2 \pi t);$ and
	(ii) $\beta(t) = 0.4t+0.5, \gamma(t) = \sqrt{t}.$ 
	
	\subsection{Comparison of one-step and two-step methods}
	We compare two methods: 
	one-step kernel weighting approach 
	and two-step approach with centering in the first step and kernel weighting in the second step. Summary tables of two-step approach using non-parametric intercept and omitting the asynchronous longitudinal covariate at the first step and kernel weighting in the second step can be found in the Appendix. This method produces results similar to the two-step approach with centering in the first step and kernel weighting at the second step. 
	
	As suggested in \citet{fan1996}, we use the Epanechnikov kernel $K(t) = 0.75(1-t^2)_{+},$ where $x_+ = \mbox{max}\{x, 0\},$ due to its excellent empirical performance. For the one-step kernel weighting method, we need kernel functions of two dimensions, where we use the product of a one-dimensional Epanechnikov kernel $K(t, s) = 0.5625(1-t^2)_{+} (1-s^2)_{+}.$ Other kernel functions produce similar results and the details are omitted.
	
	We present the results of setting (i). The results of setting (ii) can be found in the Appendix. The simulation results for $n=400$ and $n=900$ are summarized in Table \ref{tab1}. We assess the estimation and inference of $\gamma(t)$ and $\beta(t)$ at $t = 0.3, 0.6$ and $0.9.$ The results at other time points are similar and thus omitted. 
	\begin{table}[!ht]
		\caption{\label{tab1}1000 simulation results for $\beta(t)=3(t-0.4)^2$ and $\gamma(t)=\sin(2 \pi t)$}
		\centering
		\scalebox{0.6}{
			\begin{threeparttable}
				\begin{tabular}{llrrrrrrrrrrrrrr}
					\hline  
					\hline  
					&&\multicolumn{4}{c}{$t=0.3$}&&\multicolumn{4}{c}{$t=0.6$}&&\multicolumn{4}{c}{$t=0.9$}\\
					\cmidrule{3-6}\cmidrule{8-11}\cmidrule{13-16}
					BD&NP-F&Bias&SD&SE&CP&&Bias&SD&SE&CP&&Bias&SD&SE&CP\\
					\hline 
					{ $n = 400$}&&\multicolumn{14}{c}{Two-step \ (Centering+KW)}\\
					$h = n^{-0.6}, h_1 = h_2 = n^{-0.5}$&$\beta (t)$&$0.007$&$0.154$&$0.144$&$92.4$&&$0.003$&$0.130$&$0.123$&$92.5$&&$0.005$&$0.131$&$0.122$&$91.8$\\
					$h = n^{-0.7}, h_1 = h_2 = n^{-0.5}$&$\beta (t)$&$-0.008$&$0.195$&$0.186$&$92.0$&&$-0.008$&$0.166$&$0.154$&$91.2$&&$0.004$&$0.168$&$0.155$&$91.4$\\
					auto&$\beta (t)$&$-0.004$&$0.166$&$0.154$&$91.7$&&$-0.005$&$0.130$&$0.121$&$92.2$&&$0.006$&$0.137$&$0.132$&$92.6$\\
					$h = n^{-0.6}, h_1 = h_2 = n^{-0.5}$&$\gamma (t)$&$-0.047$&$0.146$&$0.139$&$90.3$&&$0.035$&$0.150$&$0.141$&$91.0$&&$0.030$&$0.148$&$0.136$&$90.2$\\
					$h = n^{-0.7}, h_1 = h_2 = n^{-0.5}$&$\gamma (t)$&$-0.046$&$0.157$&$0.157$&$91.4$&&$0.024$&$0.156$&$0.146$&$90.3$&&$0.025$&$0.149$&$0.141$&$90.8$\\
					auto&$\gamma (t)$&$-0.047$&$0.154$&$0.142$&$90.2$&&$0.023$&$0.156$&$0.140$&$89.4$&&$0.026$&$0.144$&$0.136$&$90.3$\\
					&&\multicolumn{14}{c}{One-step \ (KW)}\\
					$h_1 = h_2 = n^{-0.45}$&$\beta (t)$&$0.003$&$0.120$&$0.114$&$94.2$&&$0.002$&$0.125$&$0.116$&$91.1$&&$0.002$&$0.125$&$0.117$&$90.9$\\
					$h_1 = h_2 = n^{-0.5}$&$\beta (t)$&$0.002$&$0.148$&$0.140$&$92.0$&&$0.004$&$0.150$&$0.138$&$90.7$&&$-0.007$&$0.154$&$0.137$&$90.0$\\
					auto&$\beta (t)$&$0.001$&$0.111$&$0.106$&$92.6$&&$0.003$&$0.118$&$0.110$&$92.0$&&$0.007$&$0.114$&$0.107$&$92.4$\\
					$h_1 = h_2 = n^{-0.45}$&$\gamma (t)$&$-0.050$&$0.128$&$0.117$&$87.9$&&$0.033$&$0.121$&$0.116$&$91.1$&&$0.031$&$0.113$&$0.110$&$92.3$\\
					$h_1 = h_2 = n^{-0.5}$&$\gamma (t)$&$-0.042$&$0.156$&$0.142$&$88.2$&&$0.019$&$0.152$&$0.138$&$89.8$&&$0.019$&$0.149$&$0.130$&$87.9$\\
					auto&$\gamma (t)$&$-0.056$&$0.119$&$0.107$&$87.4$&&$0.034$&$0.115$&$0.111$&$90.8$&&$0.036$&$0.108$&$0.102$&$91.5$\\
					&\\
					{ $n = 900$}&&\multicolumn{14}{c}{Two-step \ (Centering+KW)}\\
					$h = n^{-0.6}, h_1 = h_2 = n^{-0.5}$&$\beta (t)$&$0.004$&$0.122$&$0.119$&$92.8$&&$0.001$&$0.105$&$0.100$&$92.2$&&$0.004$&$0.103$&$0.100$&$95.0$\\
					$h = n^{-0.7}, h_1 = h_2 = n^{-0.5}$&$\beta (t)$&$-0.008$&$0.176$&$0.163$&$92.7$&&$0.004$&$0.140$&$0.135$&$92.8$&&$-0.007$&$0.143$&$0.136$&$92.0$\\
					auto&$\beta (t)$&$-0.002$&$0.124$&$0.119$&$93.3$&&$0.001$&$0.105$&$0.102$&$93.0$&&$0.002$&$0.102$&$0.100$&$93.6$\\
					$h = n^{-0.6}, h_1 = h_2 = n^{-0.5}$&$\gamma (t)$&$-0.036$&$0.131$&$0.132$&$92.2$&&$0.028$&$0.134$&$0.126$&$91.8$&&$0.009$&$0.128$&$0.124$&$92.8$\\
					$h = n^{-0.7}, h_1 = h_2 = n^{-0.5}$&$\gamma (t)$&$-0.039$&$0.151$&$0.150$&$91.0$&&$0.016$&$0.141$&$0.136$&$93.0$&&$0.017$&$0.132$&$0.138$&$92.2$\\
					auto&$\gamma (t)$&$-0.031$&$0.125$&$0.121$&$91.8$&&$0.023$&$0.134$&$0.128$&$92.4$&&$0.016$&$0.135$&$0.124$&$91.2$\\
					&&\multicolumn{14}{c}{One-step \ (KW)}\\
					$h_1 = h_2 = n^{-0.45}$&$\beta (t)$&$0.000$&$0.105$&$0.102$&$92.9$&&$0.001$&$0.102$&$0.100$&$92.4$&&$-0.002$&$0.107$&$0.100$&$92.8$\\
					$h_1 = h_2 = n^{-0.5}$&$\beta (t)$&$0.002$&$0.141$&$0.126$&$91.3$&&$0.000$&$0.127$&$0.125$&$92.9$&&$-0.001$&$0.132$&$0.124$&$92.1$\\
					auto&$\beta (t)$&$0.005$&$0.092$&$0.090$&$92.7$&&$0.003$&$0.091$&$0.085$&$92.3$&&$0.001$&$0.086$&$0.083$&$93.2$\\
					$h_1 = h_2 = n^{-0.45}$&$\gamma (t)$&$-0.032$&$0.105$&$0.101$&$92.0$&&$0.014$&$0.107$&$0.100$&$92.8$&&$0.023$&$0.097$&$0.096$&$92.8$\\
					$h_1 = h_2 = n^{-0.5}$&$\gamma (t)$&$-0.022$&$0.137$&$0.125$&$92.1$&&$0.016$&$0.137$&$0.126$&$92.0$&&$0.016$&$0.128$&$0.118$&$92.5$\\
					auto&$\gamma (t)$&$-0.036$&$0.092$&$0.090$&$91.4$&&$0.026$&$0.091$&$0.086$&$92.2$&&$0.025$&$0.080$&$0.080$&$91.6$\\
					\hline 
				\end{tabular}
				\begin{tablenotes}
					\footnotesize
					\item Note: ``BD" represents the bandwidths, where $h$ represents the bandwidth in the centering approach, $h_1$ and $h_2$ represent the bandwidths in the kernel weighting (KW) approach. ``NP-F" represents the non-parametric function. ``Bias" is the absolute bias. ``SD" is the sample standard deviation. ``SE" is the average of the standard error, and ``CP" is the pointwise $95\%$ coverage probability. ``auto" represents automatic bandwidth selection.
				\end{tablenotes}
		\end{threeparttable}}
	\end{table}
	
	For the estimation of $\gamma(t)$ based on the one-step kernel weighting method or the two-step method with centering in the first step and kernel weighting in the second step, we find that the bias is generally small, the estimated standard deviation is close to the empirical standard deviation, and the probability of coverage is close to the nominal $95\%$ with a large sample size at the examined time points $t = 0.3, 0.6$ and $0.9$. Performance improves as the sample size increases. Therefore, we can conduct a valid pointwise inference of $\gamma(t)$ in practice using the proposed method. 
	
\begin{figure}
		\centering
		\includegraphics[width=16cm,height=8cm]{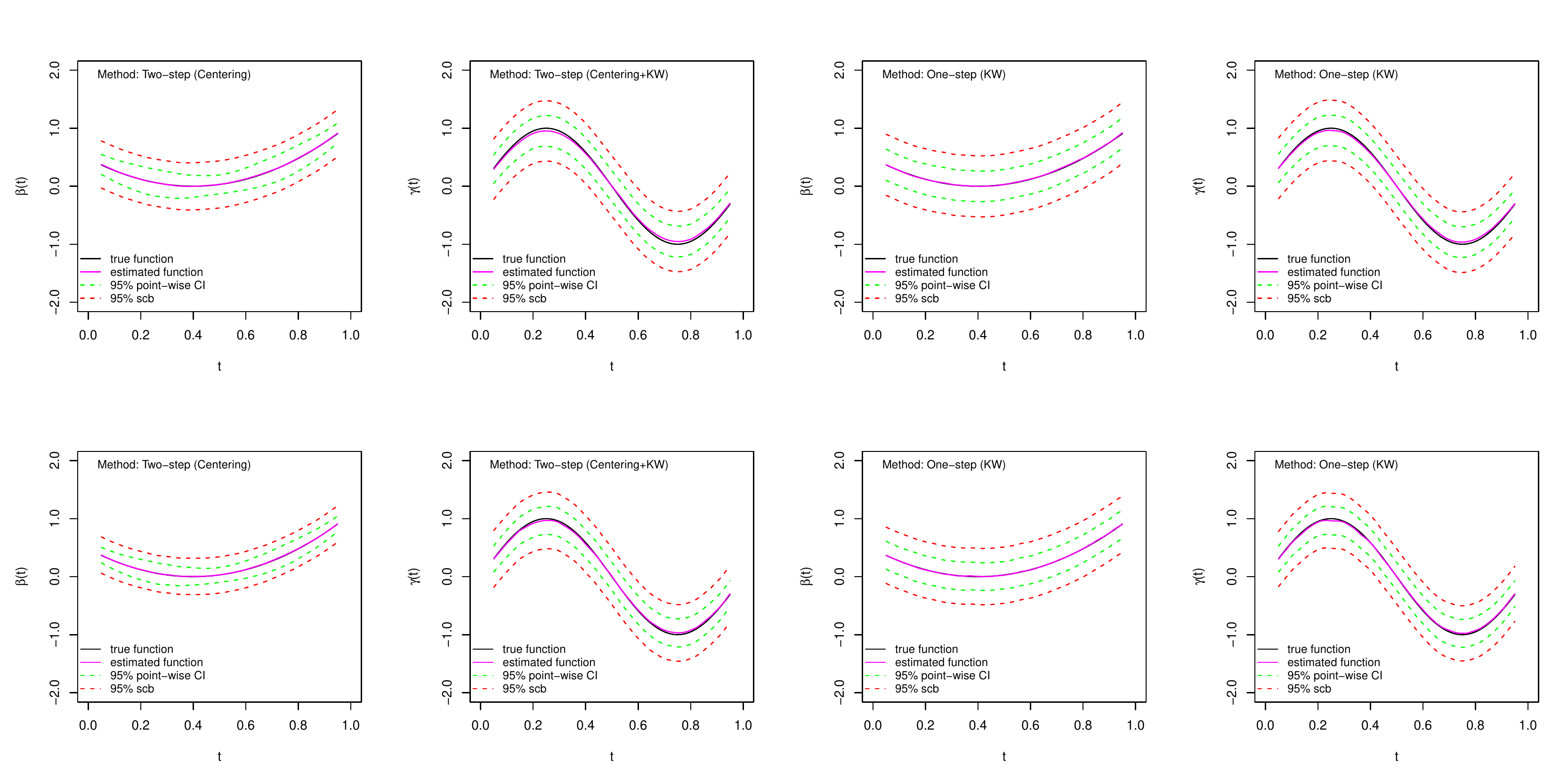}
		\caption{\label{figure2}Typical estimated curves, 95\% point-wise confidence intervals, and 95\% simultaneous confidence bands for $\beta (t)=3(t-0.4)^2, \gamma (t)=\sin(2 \pi t)$ with the same bandwidth $h=h_1=h_2=n^{-0.5},$ $n=400$ (first row) and $n=900$ (second row).}
	\end{figure}

	Both methods are valid for inference of $\beta(t)$. From Figure \ref{figure2}, we observe that with the same bandwidth, the standard deviation of $\hat{\beta}(t)$ is smaller using the two-step method with kernel weighting in the second step compared to the one-step kernel weighting method, which is consistent with our theoretical prediction of the convergence rate.  
	
	In the one-step kernel weighting method, we let $h_1 = h_2 = h.$ The allowable range of $h$ is $(n^{-1/2}, n^{-1/6}).$ In the simulation, we choose the optimal bandwidth in the range of $(n^{-0.5}, n^{-0.4}).$ In the two-step method, the allowable bandwidth in the first step for the estimation of $\beta(t)$ is $(n^{-1}, n^{-1/5})$ and we choose the optimal bandwidth in the range of $(n^{-0.8}, n^{-0.6}).$ The allowable bandwidth in the second step for the estimation of $\gamma(t)$ is $(n^{-1/2}, n^{-1/6})$ when $h_1 = h_2 = h$ and the optimal bandwidth is selected in the range $(n^{-0.5}, n^{-0.4}).$ Automatic bandwidth selection performs well and can be used to conduct inference. 

	\subsection{Simultaneous confidence bands}
	In this section, we examine the performance of the simultaneous confidence band. Instead of a point-wise confidence interval for fixed time points, it is more informative to examine the overall magnitude of variation of the non-parametric regression function. The data generation process is the same as that in Section \ref{3.1sec} and we use setting (i) to illustrate. We are interested in constructing simultaneous confidence bands for $\beta(t)$ and $\gamma(t)$ based on the one-step kernel weighting method and the two-step method with kernel weighting at the second stage. 
	
	Specifically, we use the wild bootstrap approach described in Section \ref{2.sec} to find the empirical $95$ percentile of the supremum of the difference between estimated non-parametric regression function and true non-parametric regression function. We use the Rademacher distribution, which takes value $+1$ with probability $50\%$ and value $-1$ with probability $50\%$ in the wild bootstrap. In the calculation of the empirical coverage probability, we use $181$ equally spaced grid points in $[0.05, 0.95].$ We use the Epanechikov kernel. Results based on other commonly used kernel functions such as the uniform kernel are similar, and thus omitted.
	
	In addition to the simultaneous coverage probability of the pointwise confidence interval and the newly proposed simultaneous confidence band, we also present the root average squared error (RASE). Using $\gamma(t)$ as an example, the RASE for a non-parametric function is defined as ${\rm RASE}^2 = n_{grid}^{-1} \sum\limits_{k=1}^{n_{grid}} \left\{ \hat{\gamma}(t_k) - \gamma(t_k) \right\}^2,$ where $\{ t_k, k = 1, \ldots, n_{grid} \}$ are grid points at which the non-parametric function $\gamma(\cdot)$ is estimated.
	
	The results are summarized in Table \ref{tab2} with $n=400$ and $900.$ We can see that as the sample size increases, the simultaneous confidence band coverage probabilities based on the proposed one-step kernel weighting and two-step method with kernel weighting in the second step are close to the nominal ones, and the point-wise confidence interval is not valid for simultaneous inference. The RASE gets smaller with increased sample size. 
	\begin{table}[!htpb]
		\caption{\label{tab2}Simulation results for $\beta(t)=3(t-0.4)^2, \gamma(t)=\sin(2 \pi t)$}
		\centering
		\scalebox{0.8}{
			\begin{threeparttable}
				\begin{tabular}{lrrrrrrrr}
					\hline 
					\hline  %
					&&\multicolumn{3}{c}{$n = 400$}&&\multicolumn{3}{c}{$n = 900$}\\
					\cmidrule{3-5}\cmidrule{7-9}
					BD&NP-F&RASE(SD)&CI&SCB&&RASE(SD)&CI&SCB\\
					\hline 
					&&\multicolumn{7}{c}{Two-step (Centering+KW)}\\
					$h = n^{-0.6}, h_1 = h_2 = n^{-0.5}$&$\beta (t)$&$0.134(0.022)$&$4.8$&$95.1$&&$0.109(0.014)$&$1.0$&$94.9$\\
					$h = n^{-0.7}, h_1 = h_2 = n^{-0.5}$&$\beta (t)$&$0.177(0.022)$&$0.0$&$93.6$&&$0.152(0.013)$&$0.0$&$93.8$\\
					$h = n^{-0.6}, h_1 = h_2 = n^{-0.5}$&$\gamma (t)$&$0.150(0.027)$&$3.4$&$93.1$&&$0.133(0.020)$&$1.6$&$94.1$\\
					$h = n^{-0.7}, h_1 = h_2 = n^{-0.5}$&$\gamma (t)$&$0.158(0.027)$&$1.5$&$95.4$&&$0.142(0.018)$&$0.2$&$95.0$\\
					&&\multicolumn{7}{c}{One-step (KW)}\\
					$h_1 = h_2 = n^{-0.45}$&$\beta (t)$&$0.120(0.025)$&$14.2$&$97.3$&&$0.104(0.018)$&$5.8$&$95.8$\\
					$h_1 = h_2 = n^{-0.5}$&$\beta (t)$&$0.148(0.027)$&$4.6$&$95.8$&&$0.133(0.019)$&$1.7$&$95.5$\\
					$h_1 = h_2 = n^{-0.45}$&$\gamma (t)$&$0.125(0.027)$&$9.7$&$94.8$&&$0.103(0.019)$&$7.1$&$95.2$\\
					$h_1 = h_2 = n^{-0.5}$&$\gamma (t)$&$0.149(0.026)$&$3.3$&$94.6$&&$0.133(0.020)$&$1.2$&$94.2$\\
					\hline 
				\end{tabular}\begin{tablenotes}
					\footnotesize
					\item Note: ``BD" represents the bandwidths, where $h$ represents the bandwidth in the centering approach, $h_1$ and $h_2$ represent the bandwidths in the kernel weighting (KW) approach. ``NP-F" represents the non-parametric function. ``RASE" represents the average of RASE, and ``SD" represents the standard deviation of RASE. ``CI" represents the simultaneous coverage probability based on the pointwise confidence interval. ``SCB" represents the simultaneous coverage probability based on SCB. The nominal level is $95\%$.
				\end{tablenotes}
		\end{threeparttable}}
	\end{table}
	
	Figure \ref{figure2} shows typical graphs of $\beta(t)$ and $\gamma(t)$ based on the one-step kernel weighting method and the two-step method with kernel weighting at the second step for $n=400$ and $900.$ We can see that in the estimation of $\beta(t),$ both the pointwise confidence interval and the simultaneous confidence band are narrower in the two-step method with kernel weighting at the second step than in that based on the one-step kernel weighting method. This reflects that we obtain a more efficient estimate of $\beta(t)$ by using the two-step method compared to the one-step method. As the sample size increases, we obtain narrower confidence bands for $\beta(t)$ and $\gamma(t).$.

	\section{Application to the ADNI Data}\label{4sec}
\begin{figure}[!ht]
  \centering
  \includegraphics[width=16cm,height=10cm]{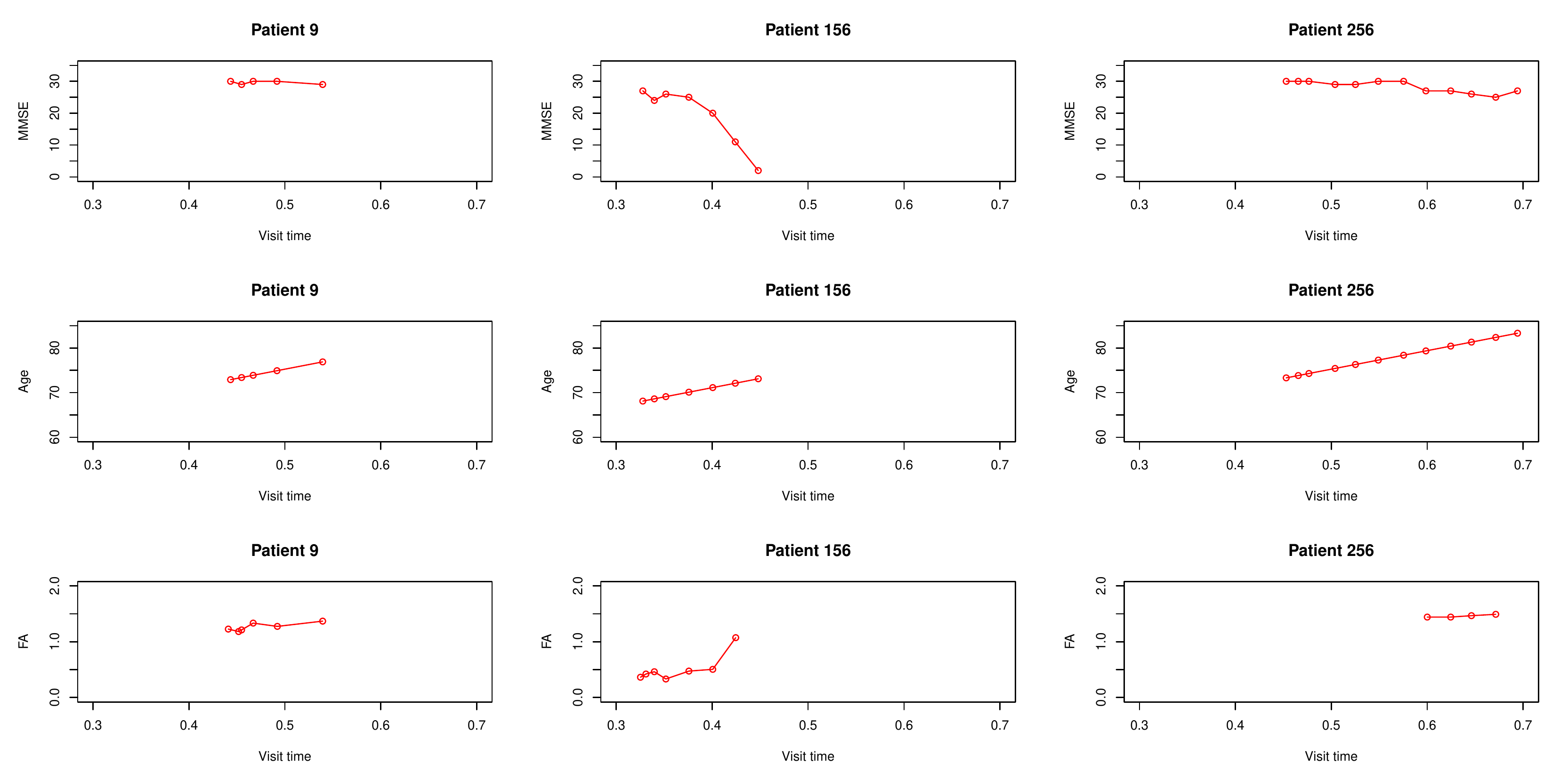}
  \caption{\label{figurer3}The growth curves for $Y_{\rm MMSE}(t), X_{\rm Age}(t)$ and $Z_{\rm FA}(t)$ of patient 9, 156 and 256.}
	\end{figure}

	In this section, we illustrate the proposed methods on a data set from the Alzheimer's Disease Neuroimaging Initiative (ADNI) study (\texttt{http://www.adni-info.org/}). ADNI is a large-scale multicenter neuroimaging study that collects genetic, clinical, imaging, and cognitive data at multiple time points to better understand Alzheimer's Disease. We used a subset of ADNI consisting of $n=256$ subjects over a $5$ year follow-up to examine the relationship between diffusion-weighted imaging (DWI) and cognitive decline. For the measurement of DWI, we use fractional anisotropy (FA), which reflects fiber density, axonal diameter, and myelination in white matter. FA is observed at $1$ to $8$ time points. For the outcome related to cognitive decline, we use the Mini-Mental State Examination (MMSE) score, with lower scores indicating impairment. The MMSE score is examined from $1$ to $13$ time points. For each subject, the measurement times of MMSE and FA are mismatched. Other covariates such as age (ranging from $55$ to 96 years), years of education (ranging from $11$ to $20$) and the number of APOE4 genes are synchronous with MMSE, creating mixed synchronous and asynchronous longitudinal covariates. The growth curves for $Y_{\rm MMSE}(t), X_{\rm Age}(t)$ and $Z_{\rm FA}(t)$ of randomly selected patients are presented in Figure \ref{figurer3}. Details of the registration and normalization of imaging data can be found in \citet{li2020}. 

	We aim to examine the association between MMSE and log hazard function of FA at region $0.35$ (the whole brain is divided into $100$ regions) accounting for other baseline covariates. We first normalize the data as recommended in \citet{diggle2002, fan2004}. Specifically, for continuous baseline covariate education level, we subtract the mean and divide the standard deviation across all individuals. For continuous longitudinal data, such as age, at time point $t$, we get Nadaraya-Watson type of estimator of $\bar{X}_{\rm {Age}}(t)$ and $\bar{X}^2_{\rm {Age}}(t).$ The standard deviation $s_{\rm {Age}}(t) = \sqrt{\bar{X}^2_{\rm {Age}}(t) - \{\bar{X}_{\rm {Age}}(t) \}^2}$ and the normalized version is $\{X_{\rm {Age}}(t) - \bar{X}_{\rm {Age}}(t)\}/s_{\rm {Age}}(t).$ Other longitudinal data can be normalized in the same way and we omit the details. Below we use the same notation for normalized data.
	
	Our model is 
	\begin{eqnarray}{\label{realdata1}}
		&&	E\{	Y_{\rm{MMSE}}(t) \mid X_{\rm{Age}}(t), X_{\rm{Edu}}, X_{\rm{APOE4}}, Z_{\rm{FA}}(t)\}  \nonumber \\
		&=& \alpha(t) + \beta_{\rm{Age}}(t) X_{\rm{Age}}(t) + \beta_{\rm {Edu}}(t) X_{\rm{Edu}} + \beta_{\rm {APOE4}}(t) X_{\rm{APOE4}} + \gamma_{\rm{FA}}(t) Z_{\rm{FA}}(t), 
	\end{eqnarray}
	where $Y_{\rm{MMSE}}(t)$ denotes longitudinal response of MMSE, $X_{\rm{Age}}(t), X_{\rm{Edu}}, X_{\rm{APOE4}}$ denote synchronous longitudinal covariates and $Z_{\rm{FA}}(t)$ denotes asynchronous longitudinal covariate. In (\ref{realdata1}), except for $X_{\rm{APOE4}},$ which denotes number of copies of gene APOE4 ($0, 1$ or $2$), the rest are continuous variables.
	
	In our proposed approaches, the two-step method requires that the synchronous and asynchronous longitudinal covariates are uncorrelated. We regress log hazard function of FA on age, education level, and the number of APOE4 genes with three univariate regression models and one multivariate regression model as follows 
	\begin{equation}{\label{realdata2}}
		E\{ Z_{\rm{FA}}(t) \mid X_{\rm{Age}}(t), X_{\rm{Edu}}, X_{\rm{APOE4}}\} = \alpha + \beta_1 X_{\rm{Age}}(t) + \beta_2 X_{\rm{Edu}} + \beta_3 X_{\rm{APOE4}}.
	\end{equation}
	The results are summarized in Table \ref{tab3}. We observe that education level and APOE4 genes are not statistically significantly associated with FA. There is a significant effect of age. We shall use one-step method to estimate regression functions in (\ref{realdata1}) and do the two-step method without age as a sensitivity analysis. The first stage of the two-step approach uses the bandwidth $h=4n^{-0.6} \approx 0.144$, and the second stage of the two-step approach and one-step kernel weighting approach use the bandwidth $h_1=h_2=4n^{-0.6} \approx 0.144$ in ADNI data.
	\begin{table}[!h]
		\caption{\label{tab3}Regression results of FA on age, education level and APOE4}
		\centering
		{\tabcolsep0.05in
			\begin{tabular}{lrrrrrrr}
				\hline
				\hline  %
				&\multicolumn{3}{c}{Fit separately}&&\multicolumn{3}{c}{Fit in one model}\\
				&Age&Edu&APOE4&&Age&Edu&APOE4\\
				\hline
				Estimate&$0.008$&$-0.001$&$-0.021$&&$0.009$&$0.001$&$0.004$\\
				SE&$0.003$&$0.007$&$0.029$&&$0.003$&$0.007$&$0.031$\\
				$p$-value&$0.003$&$0.923$&$0.475$&&$0.003$&$0.925$&$0.904$\\\hline
		\end{tabular}}
	\end{table}
	
\begin{figure}[!ht]
		\centering
		\includegraphics[width=16cm,height=8cm]{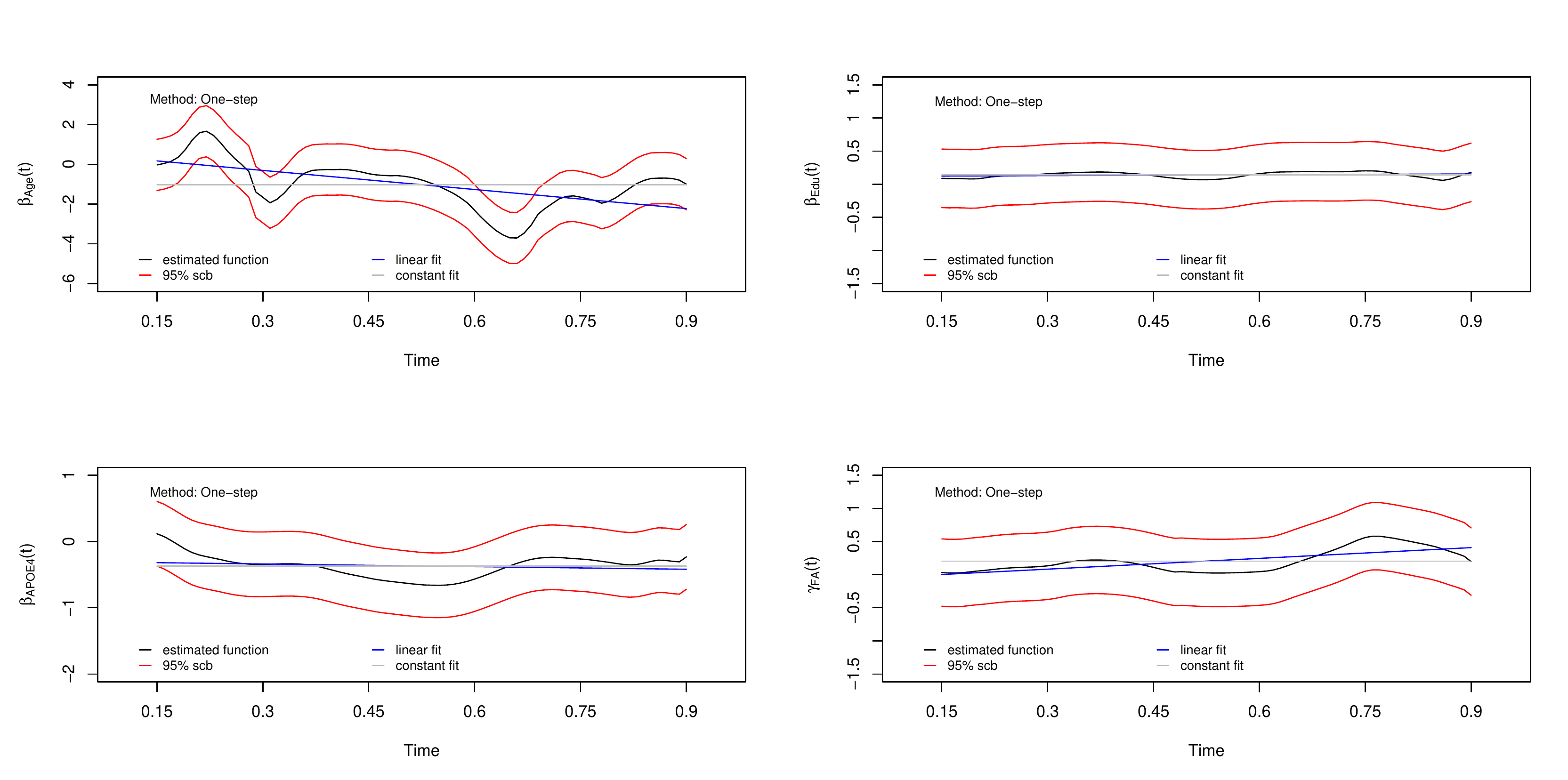}
		\caption{\label{figure3}Estimates of $\beta_{\rm {Age}}(t), \beta_{\rm {Edu}}(t), \beta_{\rm {APOE4}}(t)$ and $\gamma_{\rm {FA}}(t)$ and their SCBs in model (\ref{realdata1}). }
	\end{figure}

	Figure \ref{figure3} presents the non-parametric regression functions and SCBs of $\beta_{\rm {Age}}(t), \beta_{\rm {Edu}}(t),$ $\beta_{\rm {APOE4}}(t)$ and $\gamma_{\rm {FA}}(t)$ based on one-step method. We also draw constant and linear lines to check their adequacy in explaining the variation. We observe that the SCB of $\beta_{\rm Age}(t)$ does not contain linear or constant lines, which implies that age has a non-linear time varying effect on MMSE. Education level is not statistically significantly associated with MMSE. There is a negative constant effect of APOE4 on MMSE, which is consistent with results in the medical literature \citep{safieh2019}. FA has a non-zero constant effect on MMSE.
	Therefore our analysis suggests the following model
	\begin{eqnarray}{\label{realdata3}}
		&&	E\{	Y_{\rm{MMSE}}(t) \mid X_{\rm{Age}}(t), X_{\rm{APOE4}}, Z_{\rm{FA}}(t)\}  \nonumber \\
		&=& \alpha(t) + \beta_{\rm{Age}}(t) X_{\rm{Age}}(t) + \beta_{\rm {APOE4}} X_{\rm{APOE4}} + \gamma_{\rm{FA}} Z_{\rm{FA}}(t),
	\end{eqnarray}
	we fit model (\ref{realdata3}) based on one step method. As a comparison, we omit age and use the two step approach with centering at the first step and kernel weighting at the second step to estimate the constant coefficients for APOE4 and FA. 
	In other words, we minimize $\sum_{i=1}^{256} \sum_{j=1}^{L_i} \{ \hat{Y}_{\rm{MMSE}}(T_{ij}) - \beta_{\rm{APOE4}}\hat{X}_{\rm{APOE4}} \}^2$ to obtain $\hat{\beta}_{\rm{APOE4}}$ at the first step where $\hat{Y}_{\rm{MMSE}}$ and $\hat{X}_{\rm{APOE4}}$ are the centering estimators. Then we combine profile least squares method \citep{fan2004} and kernel weighting method \citep{cao2015} 
	to estimate $\gamma_{\rm{FA}}.$
	
\begin{figure}[!ht]
		\centering
		\includegraphics[width=8cm,height=7cm]{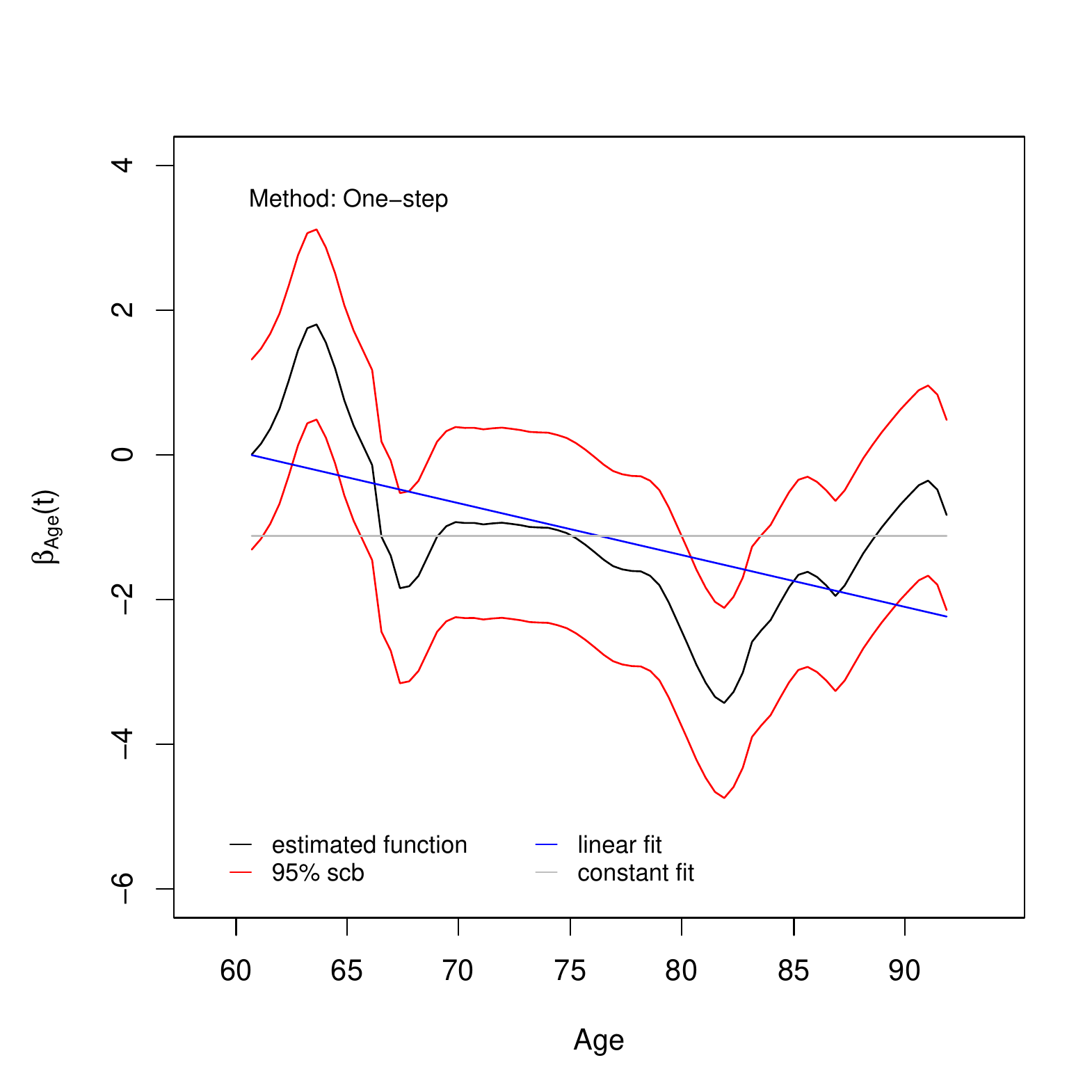}
		\caption{\label{figure4} The estimate of $\beta_{\rm Age}(t)$ and its 95\% SCB in model (\ref{realdata3}). }
	\end{figure}

	The results of constant coefficient estimates of $\beta_{\rm APOE4}$ and $\gamma_{\rm FA}$ are summarized in Table \ref{tab4}. 
	\begin{table}[!h]
		\caption{\label{tab4}Constant coefficient estimates of APOE4 and FA}
		\centering
		{\tabcolsep0.05in
			\begin{tabular}{llrrr}
				\hline\hline
				&&Estimate&SE&p-value\\
				\hline  %
				One-step&APOE4&$-0.472$&$0.078$&$<0.001$\\
				&FA&$0.129$&$0.048$&$0.008$\\
				Two-step&APOE4&$-0.471$&$0.080$&$<0.001$\\
				&FA&$0.121$&$0.050$&$0.015$\\
				\hline
		\end{tabular}}
	\end{table}
	We can see that APOE4 has a negative statistically significant association with MMSE and FA has a positive statistically significant association with MMSE based on both one-step and two-step methods. In addition, Figure \ref{figure4} presents the regression function estimation of $\beta_{\rm {Age}}(t)$ based on one-step method. It seems that the age effect is highly non-linear, which cannot be obtained by existing methods.
	
	\section{Concluding remarks}
	In this paper, we propose new methods to conduct statistical inference of mixed synchronous and asynchronous longitudinal covariates with varying coefficient models. Our one-step method allows arbitrary dependence between the synchronous and asynchronous longitudinal covariates. If we further restrict the synchronous and asynchronous longitudinal covariates to be uncorrelated, we get non-parametric rate of convergence $O(n^{2/5})$ for the regression function of synchronous longitudinal covariates, faster than the rate of convergence $O(n^{1/3})$ for the regression function of asynchronous longitudinal covariates. It remains an open question whether $O(n^{2/5})$ rate of convergence can be obtained for the regression function of asynchronous longitudinal covariates.    
	To quantify the overall magnitude of variation of the non-parametric regression function, we construct simultaneous confidence bands for the non-parametric regression functions with wild bootstrap. The computation is scalable as we put random variation in the residual without the need to compute the non-parametric regression function for each bootstrap sample. Such SCBs allow us to see if the regression function can be simplified as a constant, or a linear function.
	
	We use working independence covariance matrix similar to the GEE \citep{liangzeger86}. 
	\cite{pepe1994} shows that with time dependent covariats, working independence is a safe choice. Efficiency can be improved by correctly specified covariance matrix, which is usually difficult. 
	
	R code for simulation and data analysis can be found in \url{https://github.com/hongyuan-cao/mixed-longitudinal}. The dataset can be found in \url{https://github.com/BIG-S2/GFPLVCM}.
		\section*{Acknowledgements}
	We thank Ting Li for help with data acquisition. 
	Data used in preparation of this article were obtained from the Alzheimers Disease Neuroimaging Initiative (ADNI) database (adni.loni.usc.edu). As such, the investigators within the ADNI contributed to the design and implementation of ADNI and/or provided data but did not participate in analysis or writing of this report. A complete listing of ADNI investigators can be found at: \url{http://adni.loni.usc. edu/wp-content/uploads/how_to_apply/ADNI_Acknowledgement_List.pdf.}

	\newpage
	\appendix
	\section{Proofs of main results}
	
	In this section, we provide detailed proofs of Theorems \ref{thm1} to \ref{thm4}. Our main tools are empirical processes.
	
	\subsection{Proof of Theorem 1}
	Theorem 1. Under conditions (A1)-(A4), as $n \rightarrow \infty,$ we have  
		$$\sqrt{nh_1h_2} \left[ \{ \hat{\beta}_w(t) - \beta_0(t) \}^T, \{ \hat{\gamma}_w(t) - \gamma_0(t) \}^T \right]^T \stackrel{d}{\to} N \{ 0, A^{*}(t)^{-1} \Sigma^{*}(t) A^{*}(t)^{-1} \}$$
		where $$A^{*}(t) = E \{ C(t, t) C(t,t)^T \} \eta(t, t)$$ and $$\Sigma^{*}(t) = \big\{\iint K(z_1, z_2)^2 dz_1 dz_2\big\} E\left[ C(t, t) C(t, t)^T \sigma \{ t, X(t), Z(t) \}^2\right] \eta(t, t).$$ 
	
	\begin{proof} 
		For any time point $t$, denote true regression function
		$$\rho_0(t) = \{ \beta_0(t)^T, \dot{\beta}_0(t)^T, \gamma_0(t)^T, \dot{\gamma}_0(t)^T \}^T,$$
		estimated regression function
		$$\hat{\rho}_w(t) = \{ \hat{\beta}_{w}(t)^T, \hat{\dot{\beta}}_{w}(t)^T, \hat{\gamma}_{w}(t)^T, \hat{\dot{\gamma}}_{w}(t)^T \}^T,$$
		and longitudinal covariates
		$$R_i(t_1, t_2, t) = \{ X_i(t_1)^T, X_i(t_1)^T (t_1 - t), Z_i(t_2)^T, Z_i(t_2)^T (t_2 - t) \}^T.$$
		Let 
		$$e^{*} = \begin{pmatrix}e_{1, p \times p}&0_{p \times p}&0_{p \times q}&0_{p \times q}\\0_{q \times p}&0_{q \times p}&e_{2, q \times q}&0_{q \times q}\end{pmatrix}^T$$
		be a matrix with dimension $(2p+2q) \times (p+q),$ where $e_{1, p \times p}$ is a $p \times p$ identity matrix, $e_{2, q \times q}$ is a $q \times q$ identity matrix and the others are matrices with $0$ entries. Then our estimating equation becomes
		\begin{align}
			& {e^{*}}^T U_{n}^{w} \left\{ \rho(t) \right\}  \notag\\
			\triangleq& n^{-1} \sum\limits_{i=1}^{n} \iint K_{h_1, h_2}(t_1 - t, t_2 - t) {e^{*}}^T R_i(t_1, t_2, t) \left\{ Y_i(t_1) - R_i(t_1, t_2, t)^T \rho(t) \right\} dN^*_i(t_1, t_2) \notag\\
			=& n^{-1} \sum\limits_{i=1}^{n} \iint K_{h_1, h_2}(t_1 - t, t_2 - t) C_i(t_1, t_2) \left\{ Y_i(t_1) - R_i(t_1, t_2, t)^T \rho(t) \right\} dN^*_i(t_1, t_2), \tag{S.1}
			\label{S.1}
		\end{align}
		where $C_i(t_1, t_2) = \{ X_i(t_1)^T, Z_i(t_2)^T \}^T, i =1, \ldots, n$. Solving $U_{n}^{w} \{ \rho(t) \} = 0,$ we obtain $\hat{\rho}_w(t)$. We first show the consistency of $\hat{\beta}_w(t)$ and $\hat{\gamma}_w(t).$ Note that
		\begin{align*}
			& E \left[{e^{*}}^T U_n^w\{\rho(t)\}\right] \\
			= & n^{-1} \sum\limits_{i=1}^n E\bigg( E \bigg[ \iint K_{h_1, h_2}(t_1-t, t_2-t) {e^{*}}^T R_i(t_1, t_2, t) \left\{ Y_i(t_1) - R_i(t_1, t_2, t)^T \rho(t) \right\} \bigg.\bigg.\\
			& \bigg.\bigg.\times dN_i^*(t_1, t_2) \mid X(t), Z(s), N^*(t, s), t, s \in [0, 1] \bigg] \bigg)\\
			= & n^{-1} \sum\limits_{i=1}^n E \bigg[ \iint K_{h_1, h_2} (t_1-t, t_2-t) {e^{*}}^T R_i(t_1, t_2, t) \bigg.\\
			& \bigg.\times E \left\{Y_i(t_1) - R_i(t_1, t_2, t)^T \rho(t) \mid X(t), Z(s), t, s \in [0, 1]\right\} \eta(t_1, t_2) dt_1 dt_2 \bigg] \\
			= & n^{-1} \sum\limits_{i=1}^n E \bigg( \iint K_{h_1, h_2} (t_1-t, t_2-t)  {e^{*}}^T R_i(t_1, t_2, t) \bigg.\\
			& \bigg.\times \left[R_i(t_1, t_1, t)^T \rho_0(t) - R_i(t_1, t_2, t)^T \rho(t) + O_p\{(t_1-t)^2\} \right] \eta(t_1, t_2) dt_1 dt_2 \bigg).
		\end{align*}
		By conditions (A2)-(A4) and Taylor expansion, 
		\begin{align*}
			E \left[ {e^{*}}^T U_n^w\{\rho(t)\} \right] = & E \left\{ C(t, t)C(t, t)^T \right\} \eta(t, t) \left[ \{ \beta_0(t) - \beta(t) \}^T, \{ \gamma_0(t) - \gamma(t) \}^T \right]^T\\
			& + o_p(1).
		\end{align*}
		By the law of large numbers, ${e^{*}}^T U_{n}^{w} \left\{ \rho (t) \right\} \overset{p}\rightarrow u^{w} \left\{ \beta(t), \gamma(t) \right\},$ as $n \to \infty,$ where $u^{w} \left\{ \beta(t), \gamma(t) \right\} \newline= E \left\{ C(t, t)C(t, t)^T \right\} \eta(t, t) \left[ \{ \beta_0(t) - \beta(t) \}^T, \{ \gamma_0(t) - \gamma(t) \}^T \right]^T,$ a column vector in ${\mathbb R}^{p+q}.$ Under condition (A2), $\left\{ \beta_0(t)^T, \gamma_0(t)^T \right\}^T$ is the unique solution to $u^{w} \left\{ \beta(t), \gamma(t) \right\} = 0.$ \newline$\{ \hat{\beta}_w(t)^T, \hat{\gamma}_w(t)^T \}^T$ solves the estimation equation ${e^{*}}^T U_{n}^{w} \left\{ \rho(t) \right\} = 0.$ By \citet{ag1982}, it follows that for any fixed $t,$ $\{ \hat{\beta}_w(t)^T, \hat{\gamma}_w(t)^T \}^T \stackrel{p}{\to} \{ \beta_0(t)^T, \gamma_0(t)^T \}^T$. 
		
		We next show the asymptotic normality of $\{ \hat{\beta}_w(t)^T, \hat{\gamma}_w(t)^T \}^T.$ Let $\mathcal{P}_n$ and $\mathcal{P}$ denote the empirical measure and the true probability measure, respectively. We have
		\begin{align}
			& (nh_1h_2)^{1/2} e^{*T} U_{n}^{w} \left\{ \rho(t) \right\} \notag\\
			= & (nh_1h_2)^{1/2} (\mathcal{P}_n - \mathcal{P}) \Big[ \iint K_{h_1, h_2}(t_1 - t, t_2 - t) {e^{*}}^T R(t_1, t_2, t) \{ Y(t_1) - R(t_1, t_2, t)^T \rho(t) \} \Big. \notag\\
			& \Big. \times dN^*(t_1, t_2) \Big] \notag\\
			& + (nh_1h_2)^{1/2} {{E}} \Big[ \iint K_{h_1, h_2}(t_1 - t, t_2 - t) {e^{*}}^T R(t_1, t_2, t) \{ Y(t_1) - R(t_1, t_2, t)^T \rho(t) \} \Big. \notag\\
			& \Big.\times dN^*(t_1, t_2) \Big] \notag\\
			= & \mathrm{\uppercase\expandafter{\romannumeral1}} + \mathrm{\uppercase\expandafter{\romannumeral2}}. \tag{S.2}
			\label{S.2}
		\end{align}
		For $\mathrm{\uppercase\expandafter{\romannumeral1}},$ we consider the class of functions
		\begin{equation*}
			\begin{split}
				& \left\{ (h_1h_2)^{1/2} \iint K_{h_1, h_2}(t_1-t, t_2-t) {e^{*}}^T R(t_1, t_2, t) \{ Y(t_1) - R(t_1, t_2, t)^T \rho (t) \}\right.  \\
				& \bigg.\times dN^*(t_1, t_2): \mid \rho(t) - \rho_0(t) \mid < \epsilon \bigg\}
			\end{split}
		\end{equation*}
		for any fixed $t$ and a given constant $\epsilon.$ Note that the functions in this class are differentiable and their corresponding first-order derivatives are bounded according to condition (A2), therefore the functions in this class are Lipschitz continuous in $\rho(t)$ at any fixed $t$ and the Lipschitz constant is uniformly bounded by $(h_1h_2)^{1/2} \|\iint K_{h_1, h_2}(t_1-t, t_2-t) {e^*}^T R(t_1, t_2, t) R(t_1, t_2, t)^T dN^*(t_1, t_2)\|_{\infty}.$ It can be shown that this is a P-Donsker class \citep{vw1996}. We obtain that for $\mid \rho(t) - \rho_0(t) \mid < M (nh_1h_2)^{-1/2}$  where $M$ is a constant, the first term in (\ref{S.2}) is equal to
		\begin{align}
			\mathrm{\uppercase\expandafter{\romannumeral1}} = & (nh_1h_2)^{1/2} (\mathcal{P}_n - \mathcal{P}) \left[ \iint K_{h_1, h_2}(t_1 - t, t_2 - t) {e^{*}}^T R(t_1, t_2, t) \{ Y(t_1) - R(t_1, t_2, t)^T \rho_0(t) \} \right. \notag\\
			& \bigg. \times dN^*(t_1, t_2) \bigg] + o_p(1) \notag\\
			= & (nh_1h_2)^{1/2} e^{*T}\left( U_{n}^{w} \{\rho_0(t)\} - E \left[U_{n}^{w} \{\rho_0(t)\}\right]\right) + o_p(1). \tag{S.3}
			\label{S.3}
		\end{align}
		For the second term on the right-hand side of equation (\ref{S.2}), we have
		\begin{align*}
			\mathrm{\uppercase\expandafter{\romannumeral2}} = & (nh_1h_2)^{1/2} E \Big[ \iint K_{h_1, h_2}(t_1 - t, t_2 - t) {e^{*}}^T R(t_1, t_2, t) \left\{ Y(t_1) - R(t_1, t_2, t)^T \rho_0(t) \right\} \Big.\\
			& \Big. \times dN^*(t_1, t_2) \Big] \\
			& - (nh_1h_2)^{1/2} E \left\{ \iint K_{h_1, h_2}(t_1 - t, t_2 - t) {e^{*}}^T R(t_1, t_2, t) R(t_1, t_2, t)^T dN^*(t_1, t_2) \right\}\\
			& \times \left\{\rho(t) - \rho_0(t)\right\}\\
			= & I_1 - I_2.
		\end{align*}
		Define
		\begin{align*}
			& \varphi_i(t) \\
			& = (h_1h_2)^{1/2} \iint K_{h_1, h_2}(t_1 - t, t_2 - t) {e^{*}}^T R_i(t_1, t_2, t) \left\{ Y_i(t_1) - R_i(t_1, t_2, t)^T \rho_0(t) \right\} dN^*_i(t_1, t_2),
		\end{align*}
		then $(nh_1h_2)^{1/2} e^{*T}U_{n}^{w} \left\{ \rho_0(t) \right\} = n^{-1/2} \sum\limits_{i=1}^n \varphi_i(t),$ and we have
		\begin{align*}
			I_1 = & \sqrt{n} E \left\{ \varphi(t) \right\} \\
			= & \sqrt{n} E \left[ E\left\{ \varphi(t) \mid X(t), Z(s), N^*(t, s), t, s \in [0, 1] \right\} \right] \\
			= & (nh_1h_2)^{1/2} E \bigg[ \iint K_{h_1, h_2} (t_1-t, t_2-t) {e^{*}}^T R(t_1, t_2, t) \bigg.\\
			& \bigg. \times E \left\{Y(t_1) - R(t_1, t_2, t)^T \rho_0 (t) \mid X(t), Z(s), t, s \in [0, 1] \right\} \eta (t_1, t_2) dt_1dt_2 \bigg]\\
			= & (nh_1h_2)^{1/2} E \bigg[ \iint K_{h_1, h_2} (t_1-t, t_2-t) {e^{*}}^T R(t_1, t_2, t) \bigg.\\
			& \bigg. \times \left\{X(t_1)^T \beta_0(t_1) + Z(t_1)^T \gamma_0(t_1) - R(t_1, t_2, t)^T \rho_0 (t) \right\} \eta (t_1, t_2) dt_1dt_2 \bigg]\\
			= & (nh_1h_2)^{1/2} \iint K_{h_1, h_2} (t_1-t, t_2-t) E \bigg[ {e^{*}}^T R(t_1, t_2, t) \left\{ R(t_1, t_1, t)^T -  R(t_1, t_2, t)^T \right\}\bigg.\\
			& \bigg. \times \rho_0 (t) + O_p\{(t_1-t)^2\} \bigg] \eta (t_1, t_2) dt_1dt_2, 
		\end{align*}
		By conditions (A3) and (A4), $I_1 = o(1).$
		
		\noindent For $I_2,$ we have
		\begin{equation*}
			\begin{split}
				I_2 = & (nh_1h_2)^{1/2} {{E}} \left\{ C(t, t) C(t, t)^T \right\} \eta (t, t) \left[ \{\beta(t) - \beta_0(t)\}^T, \{\gamma(t) - \gamma_0(t)\}^T \right]^T \\
				& + O \left\{ (nh_1h_2)^{1/2}(h_1^2+h_2^2) \right\}.
			\end{split}
		\end{equation*}
		As $n \to \infty$ and $(nh_1h_2)^{1/2} (h_1^2 + h_2^2) \to 0,$
		\begin{equation}
			\mathrm{\uppercase\expandafter{\romannumeral2}} 
			= -(nh_1h_2)^{1/2} {{E}} \left\{ C(t, t) C(t, t)^T \right\} \eta(t, t) \left[ \{\beta(t) - \beta_0(t)\}^T, \{\gamma(t) - \gamma_0(t)\}^T \right]^T + o(1). \tag{S.4}
			\label{S.4}
		\end{equation}
		Combining \eqref{S.2}, \eqref{S.3} and \eqref{S.4}, we obtain
		\begin{align}
			& (nh_1h_2)^{1/2} e^{*T}\left( U_{n}^{w} \left\{ \rho_0(t) \right\} - E \left[ U_{n}^{w} \left\{ \rho_0(t) \right\} \right]\right) \notag \\
			=&(nh_1h_2)^{1/2}{{E}} \left\{ C(t, t) C(t, t)^T \right\} \eta(t, t) \left[ \{\hat{\beta}_w(t) - \beta_0(t)\}^T, \{\hat{\gamma}_w(t) - \gamma_0(t)\}^T \right]^T + o(1).\tag{S.5}
			\label{S.5}
		\end{align}
		Recall 
		\begin{align*}
			& \varphi_i(t)\\
			& = (h_1h_2)^{1/2} \iint K_{h_1, h_2}(t_1 - t, t_2 - t) {e^{*}}^T R_i(t_1, t_2, t) \{ Y_i(t_1) - R_i(t_1, t_2, t)^T \rho_0(t) \} dN^*_i(t_1, t_2).
		\end{align*}
		Since $(nh_1h_2)^{1/2} e^{*T}U_{n}^{w} \left\{ \rho_0(t) \right\} = n^{-1/2} \sum\limits_{i=1}^n \varphi_i(t)$, sum of independent and identically distributed (i.i.d.) random variables,  
		we next calculate $\mbox{var} \left\{ \varphi(t) \right\}.$ Note that 
		\begin{align*}
		    \mbox{var} \left\{ \varphi(t) \right\} & = E \left[ \mbox{var} \left\{ \varphi(t) \mid X(t), Z(s), N^*(t, s), t, s \in [0, 1] \right\} \right] \\
		    & + \mbox{var} \left[ E \left\{ \varphi(t) \mid X (t), Z(s), N^*(t, s), t, s \in [0, 1] \right\} \right],
		\end{align*}
		where
		\begin{align*}
			& E \left[ \mbox{var} \left\{ \varphi(t) \mid X(t), Z(s), N^*(t, s), t, s \in [0, 1] \right\} \right] \\
			= & h_1h_2 E \bigg[ \iiiint K_{h_1, h_2}(t_1-t, t_2-t) K_{h_1, h_2}(s_1-t, s_2-t) {e^{*}}^T R(t_1, t_2, t) R(s_1, s_2, t)^T e^{*} \bigg.\\
			& \times \bigg. E \left\{Y(t_1) Y(s_1) \mid X(t), Z(s), N^*(t, s), t, s \in [0, 1]\right\} dN^*(t_1, t_2) dN^*(s_1, s_2)\bigg]\\
			& - h_1h_2 E \bigg[ \iiiint K_{h_1, h_2}(t_1-t, t_2-t) K_{h_1, h_2}(s_1-t, s_2-t) {e^{*}}^T R(t_1, t_2, t) R(s_1, s_2, t)^T e^{*} \bigg.\\
			& \times \bigg. E \left\{Y(t_1) \mid X(t), Z(s), N^*(t, s), t, s \in [0, 1]\right\}\bigg. \\
			& \times \bigg. E \left\{Y(s_1) \mid X(t), Z(s), N^*(t, s), t, s \in [0, 1]\right\} dN^*(t_1, t_2) dN^*(s_1, s_2)\bigg]\\
			= & \iint K(z_1, z_2)^2 dz_1 dz_2 {{E}} \left[ C(t, t) C(t, t)^T \sigma \left\{t, X(t), Z(t)\right\}^2\right] \eta(t, t) + O(h_1^2 + h_2^2),
		\end{align*}
		and
		\begin{align*}
			& \mbox{var} \left[ E\left\{ \varphi(t) \mid X(t), Z(s), N^*(t, s), t, s \in [0,1] \right\} \right] \\
			= & h_1 h_2 \mbox{var} \bigg( \iint K_{h_1, h_2}(t_1-t, t_2-t) {e^{*}}^T R(t_1, t_2, t) \left[ \{ R(t_1, t_1, t) - R(t_1, t_2, t) \}^T \rho_0(t) \right. \bigg.\\
			& \bigg.\left. + O_p \left\{ (t_1-t)^2 \right\} \right]  dN^*(t_1, t_2) \bigg) \\
			= & h_1 h_2 E \bigg( \iiiint K_{h_1, h_2}(t_1-t, t_2-t) K_{h_1, h_2}(s_1-t, s_2-t) {e^{*}}^T R(t_1, t_2, t) \bigg. \\
			& \bigg. \times \left[ \left\{ R(t_1, t_1, t) - R(t_1, t_2, t) \right\}^T \rho_0(t) + O_p \left\{ (t_1-t)^2 \right\} \right] \bigg. \\
			& \bigg. \times \left[ \left\{ R (s_1, s_1, t) - R (s_1, s_2, t) \right\}^T \rho_0 (t) + O_p \left\{ (s_1-t)^2 \right\} \right] R(s_1, s_2, t)^T e^{*} \bigg.\\
			& \bigg. \times dN^*(t_1, t_2) dN^*(s_1, s_2) \bigg) \\
			& - h_1 h_2 \bigg\{ \iint K_{h_1, h_2}(t_1-t, t_2-t) E \bigg( {e^{*}}^T R(t_1, t_2, t) [ \left\{ R(t_1, t_1, t) - R(t_1, t_2, t) \right\}^T \rho_0 (t)  \bigg.\bigg.\\
			& \bigg.\bigg. + O_p\left\{ (t_1-t)^2 \right\} ] \bigg)  dN^*(t_1, t_2) \bigg\}^{\otimes 2} \\
			= & O(h_1^2 + h_2^2).
		\end{align*}
		
		To prove the asymptotic normality, we verify the Lyapunov condition. Note that
		$$(nh_1h_2)^{1/2} e^{*T} U_{n}^{w} \left\{ \rho_0(t) \right\} = n^{-1/2} \sum\limits_{i=1}^n \varphi_i(t) = \sum\limits_{i=1}^n n^{1/2} n^{-1} \varphi_i(t),$$
		then similar to the calculation of variance, we have
		\begin{align*}
			\sum\limits_{i=1}^n E \left[\mid n^{1/2} n^{-1} \varphi_i(t) - E\{ n^{1/2} n^{-1} \varphi_i(t) \}\mid^3\right] &= n O \left\{(nh_1h_2)^{3/2} n^{-3} (h_1h_2)^{-2}\right\} \\
			&= O \left\{(nh_1h_2)^{-1/2}\right\}.
		\end{align*}
		Therefore, by condition (A4) $nh_1h_2 \to \infty$ and $(nh_1h_2)^{1/2} (h_1^2 + h_2^2) \to 0,$ we have
		$$(nh_1h_2)^{1/2} e^{*T}\left( U_n^w \left\{ \rho_0(t) \right\} - E \left[ U_n^w \left\{ \rho_0(t) \right\} \right] \right) \xrightarrow{d} N \left\{ 0, \Sigma^{*}(t) \right\},$$
		where $\Sigma^{*}(t) = \big\{\iint K (z_1, z_2)^2 dz_1 dz_2\big\} E \left[ C(t, t) C(t, t)^T \sigma \left\{t, X(t), Z(t)\right\}^2\right] \eta (t, t).$
		Combining with equation \eqref{S.5}, we have
		$$\sqrt{nh_1h_2} \left[ \{ \hat{\beta}_w(t) - \beta_0(t) \}^T, \{ \hat{\gamma}_w(t) - \gamma_0(t) \}^T \right]^T \stackrel{d}{\to} N \{ 0, A^{*}(t)^{-1} \Sigma^{*}(t) A^{*}(t)^{-1} \},$$
		where $$A^{*}(t) = E \{ C(t, t) C(t,t)^T \} \eta(t, t)$$ and $$\Sigma^{*}(t) =  \big\{ \iint K(z_1, z_2)^2 dz_1 dz_2 \big\}  E\left[ C(t, t) C(t, t)^T \sigma \{ t, X(t), Z(t) \}^2\right] \eta(t, t).$$ 
		Therefore the conclusion of Theorem $1$ holds.
	\end{proof}
	
	\subsection{Proof of Theorem 2}
	Theorem 2. Under conditions (\ref{orthogonality}), (A5)-(A8),  as $n\rightarrow \infty,$ we have 
		$$\sqrt{nh} \{\hat{\beta}_c(t) - \beta_0(t)\} \stackrel{d}{\to} N \{0, A^{-1}(t) \Sigma(t) A^{-1}(t)\},$$
		where
		\begin{equation*}
			\begin{split}
				A(t) & = E\{\tilde{X}(t) \tilde{X}(t)^T\} \lambda(t)\\
				\mbox{and} \quad \Sigma(t) & =  \big\{\int K(z)^2 dz \big\} E \left[ \tilde{X}(t) \tilde{X}(t)^T \sigma \{t, \tilde{X}(t)\}^2 \right] \lambda(t).
			\end{split}
		\end{equation*}
	
	\begin{proof}
		For any $t,$ let
		$$\theta_0(t) = \{ \beta_0(t)^T, \dot{\beta}_0(t)^T \}^{T},$$ 
		$$\hat{\theta}_c(t) = \{ \hat{\beta}_c(t)^T, \hat{\dot{\beta}}_c(t)^T \}^{T},$$ 
		and
		$$\tilde{W}_i (t_1, t) = \{ \tilde{X}_i (t_1)^T, \tilde{X}_i (t_1)^T (t_1 - t) \}^T,$$ 
		under conditions (A7) and (A8), $\hat{m}_X(t) \stackrel{p}{\to} m_X(t)$ and $\hat{m}_Y(t) \stackrel{p}{\to} m_Y(t).$ Let $e= (e_{1, p \times p}, 0_{p \times p})^T \in {\mathbb R}^{2p \times p},$ where $e_{1, p \times p}$ is a $p \times p$ unit matrix, then the estimating equation is 
		\begin{equation}
			e^T U_{n}^{c} \left\{ \theta(t) \right\} = n^{-1} \sum\limits_{i=1}^{n} \int K_h(t_1 - t) e^T \tilde{W}_i(t_1, t) \{ \tilde{Y}_i(t_1) - \tilde{W}_i(t_1, t)^T \theta(t) \} dN_i(t_1) + o_p(1). \tag{S.6}
			\label{S.6}
		\end{equation}
		Solving $e^T U_{n}^{c} \{ \theta(t) \} = 0,$ we obtain $\hat{\beta}_c(t)$. We first show the consistency of $\hat{\beta}_c(t).$ 
		\begin{equation*}
			\begin{split}
				& E \left[\int K_h(t_1 - t) e^T \tilde{W}(t_1, t) \left\{ \tilde{Y}(t_1) - \tilde{W}(t_1, t)^T \theta(t) \right\} dN(t_1)\right] \\
				= & E\left( E \left[ \int K_{h}(t_1-t) e^T \tilde{W}(t_1, t) \left\{\tilde{Y}(t_1) - \tilde{W}(t_1, t)^T \theta(t)\right\}dN(t_1) \mid X(t), N(t), t \in [0, 1] \right] \right)\\
				= & E \left[ \int K_{h} (t_1-t) e^T \tilde{W}(t_1, t) E \left\{\tilde{Y}(t_1) - \tilde{W}(t_1, t)^T \theta(t) \mid X(t), t \in [0, 1]\right\} \lambda(t_1) dt_1 \right] \\
				= & E \left( \int K_{h} (t_1-t) \left[ e^T \tilde{W}(t_1, t) \tilde{W}(t_1, t)^T \left\{ \theta_0(t) - \theta(t) \right\} + O_p\{(t_1-t)^2\} \right] \lambda(t_1) dt_1 \right).
			\end{split}
		\end{equation*}
		Let $t_1 = t+hz$, by conditions (A6)-(A8) and the law of large numbers, we have ${e}^T U_n^c \{\theta(t)\} \stackrel{p}{\to} \{\beta_0(t) - \beta(t)\} E \{\tilde{X}(t) \tilde{X}(t)^T\} \lambda(t)$ as $n \to \infty.$
		Similar to the proof of Theorem $1$, under condition (A6) and convexity lemma \citep{ag1982}, $\hat{\beta}_c(t) \stackrel{p}{\to} \beta_0(t)$.
		
		Next we show the asymptotic normality of $\hat{\beta}_c(t).$ Let $\mathcal{P}_n$ and $\mathcal{P}$ denote the empirical measure and the true probability measure respectively, we have
		\begin{align}
			\sqrt{nh} e^T U_{n}^{c} \left\{ \theta(t) \right\} = & \sqrt{nh} \left( \mathcal{P}_n - \mathcal{P} \right) \left[ \int K_h(t_1 - t) e^T \tilde{W}(t_1, t) \{ \tilde{Y}(t_1) - \tilde{W}(t_1, t)^T \theta(t) \} dN(t_1) \right] \notag\\
			& + \sqrt{nh} {{E}} \left[ \int K_h(t_1 - t) e^T \tilde{W}(t_1, t) \{ \tilde{Y}(t_1) - \tilde{W}(t_1, t)^T \theta(t) \} dN(t_1) \right].\tag{S.7} 
			\label{S.7}
		\end{align}
		For the first term in \eqref{S.7}, we consider the class of functions
		$$\left\{\sqrt{h} \int K_h(t_1 - t) e^T \tilde{W}(t_1, t) \{ \tilde{Y}(t_1) - \tilde{W}(t_1, t)^T \theta(t) \} dN(t_1): \mid \theta(t)-\theta_0(t) \mid <\epsilon \right\}$$
		for any given constant $\epsilon$ and any fixed time point $t.$ Similarly to the proof in Theorem $1$, it can be shown that this is a P-Donsker class \citep{vw1996}, we obtain that for $\mid \theta(t) - \theta_0(t) \mid < M(nh)^{-1/2}$ where $M$ is a constant, the first term in \eqref{S.7} is equal to
		\begin{align}
			& \sqrt{nh} \left( \mathcal{P}_n - \mathcal{P} \right) \left[ \int K_h(t_1 - t) e^T \tilde{W}(t_1, t) \{ \tilde{Y}(t_1) - \tilde{W}(t_1, t)^T \theta_0(t) \} dN(t_1) \right] +o_p(1) \notag\\
			= & \sqrt{nh} e^T \left( U_{n}^{c} \{\theta_0(t)\} - E \left[ U_{n}^{c} \{\theta_0(t)\} \right] \right) + o_p(1). \tag{S.8}
			\label{S.8}
		\end{align}
		For the second term on the right-hand side of equation \eqref{S.7}, we have
		\begin{align}
			& \sqrt{nh} {{E}} \Big( \int K_h(t_1 - t) e^T \tilde{W}(t_1, t) \left[ \tilde{Y}(t_1) - \tilde{W}(t_1, t)^T \left\{ \theta(t) - \theta_0(t) \right\} - \tilde{W}(t_1, t)^T \theta_0(t) \right]\Big. \notag\\
			& \Big. \times dN(t_1) \Big) \notag\\
			= & \sqrt{nh} E \left[ \int K_h(t_1 - t) e^T \tilde{W}(t_1, t) \left\{ \tilde{Y}(t_1) - \tilde{W}(t_1, t)^T \theta_0(t) \right\} dN(t_1) \right] \notag\\
			& \ - \sqrt{nh} E \left\{ \int K_h(t_1 - t) e^T \tilde{W}(t_1, t) \tilde{W}(t_1, t)^T \lambda(t_1) dt_1 \right\} \left\{ \theta(t) - \theta_0(t) \right\} \notag\\
			= & I_1 - I_2. \tag{S.9}
			\label{S.9}
		\end{align}
		For $I_1,$ let $\psi_i(t)= \sqrt{h} \int K_h(t_1 - t) e^T \tilde{W}_i(t_1, t) \{ \tilde{Y}_i(t_1) - \tilde{W}_i(t_1, t)^T \theta_0(t) \} dN_i(t_1)$. $\psi_i(t)$s are i.i.d. and $\sqrt{nh} e^T U_{n}^{c} \left\{ \theta_0(t) \right\} = n^{-1/2} \sum\limits_{i=1}^{n} \psi_i(t).$ We have
		\begin{align*}
			I_1 = & \sqrt{n} E \left\{ \psi(t) \right\} \\
			= & \sqrt{n} E \left[ E\left\{ \psi(t) \mid \tilde{X}(t), N(t), t \in [0, 1] \right\} \right] \\
			= & (nh)^{1/2} E \left[ \int K_{h} (t_1-t) e^T \tilde{W}(t_1, t) E \left\{\tilde{Y}(t_1) - \tilde{W}(t_1, t)^T \theta_0 (t) \mid \tilde{X}(t), t \in [0, 1] \right\} \lambda(t_1) dt_1 \right]\\
			= & (nh)^{1/2} E \left[ \int K_{h} (t_1-t) e^T \tilde{W}(t_1, t) \left\{\tilde{W}(t_1, t_1)^T \theta_0(t_1) - \tilde{W}(t_1, t)^T \theta_0 (t) \right\} \lambda (t_1) dt_1 \right]\\
			= & (nh)^{1/2} \int K_{h} (t_1-t) E \Big[ e^T \left\{\tilde{W}(t_1, t) \tilde{W}(t_1, t_1)^T - \tilde{W}(t_1, t) \tilde{W}(t_1, t)^T \right\} \theta_0 (t) \Big.\\
			& \Big.+ O_p\{(t_1-t)^2\} \Big] \lambda (t_1) dt_1,
		\end{align*}
		let $t_1 = h_1z_1+t,$ similar to the proof of Theorem $1$, by conditions (A7) and (A8), we have $I_1 = o(1).$
		
		\noindent For $I_2$, we have
		\begin{equation*}
			\begin{split}
				I_2 & = \sqrt{nh} e^T E \left\{ \int K(z) \tilde{W}(t+hz, t) \tilde{W}(t+hz, t)^T \lambda(t+hz) dz \right\} \left\{ \theta(t) - \theta_0(t) \right\} \\
				& = \sqrt{nh} E \{ \tilde{X}(t) \tilde{X}(t)^T \} \lambda(t) \left\{ \beta(t) - \beta_0(t) \right\} + O(n^{1/2}h^{5/2}). \\
			\end{split}
		\end{equation*}
		Consequently,
		\begin{equation}
			\sqrt{nh} E \{ \tilde{X}(t) \tilde{X}(t)^T \} \lambda(t) \{ \hat{\beta}_c(t) - \beta_0(t) \} + o(1) = \sqrt{nh} e^T \left( U_{n}^{c} \left\{ \theta_0(t) \right\} - E \left[ U_{n}^{c} \left\{ \theta_0(t) \right\} \right] \right). \tag{S.10}
			\label{S.10}
		\end{equation}
		The variance of $\sqrt{nh} e^T U_{n}^{c} \left( \theta_0 (t) \right)$ can be computed using $\psi_i (t)$ defined earlier,
		\begin{equation*}
			\begin{split}
				\mathrm{var} \left\{ \psi(t) \right\} & = E \left[ \mathrm{var} \left\{ \psi(t) \mid \tilde{X}(t), N(t), t \in [0, 1] \right\} \right] + \mathrm{var} \left[ E \left\{ \psi(t) \mid \tilde{X}(t), N(t), t \in [0, 1] \right\} \right] \\
				& = J_1 + J_2,
			\end{split}
		\end{equation*}
		where
		\begin{align*}
			J_1 = & E \left\{ E \left(\left[\psi(t) - E\left\{\psi(t) \mid \tilde{X}(t), N(t), t \in [0, 1] \right\}\right]^2 \mid \tilde{X}(t), N(t), t \in [0, 1]\right) \right\} \\
			= & h E \bigg\{ \iint K_h(t_1 - t) K_h(s_1 - t) e^T \tilde{W}(t_1, t) \tilde{W}(s_1, t)^T e \bigg.\\
			& \bigg. \times E \left( \left[ \tilde{Y} (t_1) - E \left\{ \tilde{Y}(t_1) \mid \tilde{X}(t), N(t), t \in [0, 1] \right\} \right] \right. \bigg.\\
			& \bigg. \left. \times \left[ \tilde{Y}(s_1) - E\left\{ \tilde{Y}(s_1) \mid \tilde{X}(t), N (t), t \in [0, 1] \right\} \right] \mid \tilde{X}(t), N(t), t \in [0, 1] \right) dN (t_1)dN(s_1)\bigg\} \\
			= & \int K(z)^2 dz E \left[ \tilde{X}(t) \tilde{X}(t)^T \sigma \{t, \tilde{X}(t)\}^2\right] \lambda(t) + O(h^2),
		\end{align*}
		and
		\begin{align*}
			J_2 = & h \mathrm{var} \left[\int K_h(t_1 - t) e^T \tilde{W}(t_1, t) \left\{ \tilde{W}(t_1, t_1)^T \theta_0(t_1) - \tilde{W}(t_1, t)^T \theta_0(t) \right\} dN(t_1) \right] \\
			= & h E \bigg[ \iint K_h(t_1 - t) K_h(s_1 - t) e^T \tilde{W}(t_1, t) \left\{ \tilde{W}(t_1, t_1)^T \theta_0(t_1) - \tilde{W}(t_1, t)^T \theta_0(t) \right\} \bigg.\\
			& \times \bigg.\left\{ \tilde{W}(s_1, s_1)^T \theta_0(s_1) - \tilde{W}(s_1, t)^T \theta_0(t) \right\} \tilde{W}(s_1, t)^T e dN(t_1) dN(s_1) \bigg] \\
			& - h \left\{\int K_h(t_1 - t) e^T E \left( \tilde{W}(t_1, t) \left[ \tilde{W}(t_1, t_1)^T \theta_0(t_1) - \tilde{W}(t_1, t)^T \theta_0(t) \right] \right) dN(t_1) \right\}^{\otimes 2}\\
			= & O(h^2).
		\end{align*}
		
		To prove the asymptotic normality, we verify the Lyapunov condition. Note that 
		$$(nh)^{1/2} e^T U_{n}^{c} \left\{ \theta_0(t) \right\} = n^{-1/2} \sum\limits_{i=1}^n \psi_i(t) = \sum\limits_{i=1}^n n^{1/2} n^{-1} \psi_i(t),$$ 
		then similar to the calculation of variance, 
		\begin{equation*}
			\sum\limits_{i=1}^n E \left[\mid n^{1/2} n^{-1} \psi_i(t) - E\{ n^{1/2} n^{-1} \psi_i(t) \}\mid^3\right] = n O \left\{(nh)^{3/2} n^{-3} h^{-2}\right\} = O \left\{(nh)^{-1/2}\right\}.
		\end{equation*}
		Combining \eqref{S.10}, we have
		$$\sqrt{nh} \{ \hat{\beta}_c(t) - \beta_0(t) \} \stackrel{d}{\to} N \{ 0, A(t)^{-1} \Sigma(t) A(t)^{-1} \}, $$
		where \begin{equation*}
			\begin{split}
				A(t) & = E \{ \tilde{X}(t) \tilde{X}(t)^T \} \lambda(t)\\
				\mbox{and} \quad \Sigma(t) & =  \big\{ \int K(z)^2 dz \big\}  E \left[ \tilde{X}(t) \tilde{X}(t)^T \sigma \{ t, \tilde{X}(t) \}^2\right] \lambda(t).
			\end{split}
		\end{equation*}
		Therefore the conclusion of Theorem $2$ holds.
	\end{proof}
	
	\subsection{Proof of Theorem 3}
	Theorem 3.Under conditions (\ref{orthogonality}), (A5)-(A8),   as $n \rightarrow \infty,$ we have
		$$
		\sqrt{nh} \{ \hat{\beta}_v(t) - \beta_0(t) \} \stackrel{d}{\to} N \{ 0, A^{-1}(t) \Sigma(t) A^{-1}(t)\},
		$$
		where $A(t)$ and $\Sigma(t)$ are specified in Theorem $2$.
	
	\begin{proof}
		From \citet{fan2008}, $\hat{\iota}_v(t) \triangleq \{ \hat{\alpha}_v(t), \hat{\beta}_v(t)^T, \hat{\dot{\alpha}}_v(t), \hat{\dot{\beta}}_v(t)^T \}^T$ is consistent estimates of the coefficients in model (2.8). Here we show that $\hat{\beta}_v(t)$ is a consistent estimator for $\beta(t)$ in model (2.1). For any fixed time point $t$, define
		\begin{equation}
			U_{n}^{v} \{ \iota(t) \} = n^{-1} \sum\limits_{i=1}^{n} \int K_{h}(t_1 - t) \left\{ Y_i(t_1) - Q_{i}(t_1, t)^T \iota(t) \right\}^2 dN_i(t_1), \tag{S.11}
			\label{S.11}
		\end{equation}
		where $\iota(t) = \{ \alpha(t), \beta(t)^T, \dot{\alpha}(t), \dot{\beta}(t)^T \}^T$ and $Q_i(t_1, t) = \{ 1, X_i(t_1)^{T}, (t_1 - t), X_i(t_1)^T (t_1 -t) \}^T$. From (2.5),
		$\alpha (t) = E\{Z(t)\}^T \gamma(t) = E\{Y(t)\} - E\{X(t)\}^{T} \beta(t),$ we have
		\begin{align}
			& U_{n}^{v} \{ \iota(t) \} \notag\\
			= & n^{-1} \sum\limits_{i=1}^{n} \int K_{h}(t_1 - t) \left\{ Y_i(t_1) - \alpha(t) - \dot{\alpha}(t) (t_1 - t) - X_i(t_1)^T \beta(t) - X_i(t_1)^T \dot{\beta}(t) (t_1 - t) \right\}^2 \notag\\
			& \times dN_i (t_1) \notag\\
			= & n^{-1} \sum\limits_{i=1}^{n} \int K_{h}(t_1 - t) \Big\{ Y_i(t_1) - EY(t_1) + O_p(t_1 - t)^2 + EX(t_1)^T \beta(t_1) - X_i(t_1)^T \beta(t)\Big.\notag\\
			& \Big. - X_i(t_1)^T \dot{\beta}(t) (t_1 - t) \Big\}^2 dN_i(t_1) \notag\\
			= & n^{-1} \sum\limits_{i=1}^{n} \int K_{h}(t_1 - t) \left\{ \tilde{Y}_i(t_1) - \tilde{W}_i(t_1, t)^T \theta(t) \right\}^2 dN_i(t_1) +o_p(1), \tag{S.12}
			\label{S.12}
		\end{align}
		where $\theta(t) = \{ \beta(t)^T, \dot{\beta}(t)^T \}^{T}$, $\tilde{W}_i(t_1, t) = \{ \tilde{X}_i(t_1)^T, \tilde{X}_i(t_1)^T (t_1 - t) \}^T,$ $\tilde{X}_i(t_1) = X_i(t_1) - E X(t_1),$ and $\tilde{Y}_i(t_1) = Y_i(t_1) - E Y(t_1).$ The estimating equation is the same as the centering approach. Consequently, let $\hat{\theta}_v (t) = \{ \hat{\beta}_v(t)^T, \hat{\dot{\beta}}_v(t)^T \}^{T}$ be the estimating function using the varying-coefficient model approach and $\theta_0 (t)$ be the true function. The proof of Theorem $3$ is the same as that of Theorem $2$ and is thus omitted.
	\end{proof}
	
	\subsection{Proof of Theorem 4}
	Theorem 4. 	Under conditions (\ref{orthogonality}), (A1)-(A4),  as $n\rightarrow \infty,$ we have 
 $$\sqrt{nh_1h_2} \{ \hat{\gamma}(t) - \gamma_0(t) \} \stackrel{d}{\to} N \{ 0, A^{+}(t)^{-1} \Sigma^{+}(t) A^{+}(t)^{-1} \},$$
		where 
		\begin{equation*}
			\begin{split}
				A^{+}(t) & = E \{ Z(t) Z(t)^T \} \eta(t, t)\\
				\mbox{and} \quad \Sigma^{+}(t) & = \big\{ \iint K(z_1, z_2)^2 dz_1 dz_2 \big\}  E\left[ Z(t) Z(t)^T \sigma \{ t, X(t), Z(t) \}^2\right] \eta(t, t).
			\end{split}
		\end{equation*}
	
	\begin{proof} 
		For any fixed time point $t,$ let $\phi_0(t) = \{ \gamma_0(t)^T, \dot{\gamma}_0(t)^T \}^T$, $\hat{\phi}(t) = \{ \hat{\gamma}(t)^T, \hat{\dot{\gamma}}(t)^T \}^T,$ $V_i(t_2, t) = \{ Z_i(t_2)^T, Z_i(t_2)^T (t_2 - t) \}^T,$ $\theta_0(t) = \{ \beta_0(t)^T, \dot{\beta}_0(t)^T \}^T,$ $\hat{\theta}(t) = \{ \hat{\beta}(t)^T, \hat{\dot{\beta}}(t)^T \}^T,$ and $W_i(t_1, t) = \{ X_i(t_1)^T, X_i(t_1)^T (t_1 - t) \}^T,$ for $|\hat{\theta}(t) - \theta_0(t) | < \epsilon$ with any given constant $\epsilon,$ the estimating equation about $\phi(t)$ becomes
		\begin{align}
			& U_n \{ \phi(t), \hat{\theta}(t) \} \notag\\
			= & n^{-1} \sum_{i=1}^{n} \iint K_{h_1, h_2} (t_1 - t, t_2 - t) V_i (t_2, t) \left\{ Y_i (t_1) - W_i (t_1, t)^T \theta_0 (t) -  V_i (t_2, t)^T \phi (t) \right\} \notag\\
			& \times dN^*_i (t_1, t_2) + o_p(1). \tag{S.13}
			\label{S.13}
		\end{align}
		Let $e^{+}= (e_{2, q \times q}, 0_{q \times q})^T \in {\mathbb R}^{2q \times q}$. 
		By taking expected value of ${e^{+}}^T U_{n} \{ \phi(t), \hat{\theta}(t) \}$, we have
		\begin{align*}
			& n^{-1} \sum\limits_{i=1}^n {e^{+}}^T E \Big[ \iint K_{h_1, h_2} (t_1-t, t_2-t) V_i(t_2, t) E \left\{Y_i(t_1) - W_i(t_1, t)^T \theta_0(t) \right.\Big.\\
			& \left.\Big.- V_i(t_2, t)^T \phi(t) \mid X(t), Z(s), t, s \in [0, 1]\right\} \eta(t_1, t_2) dt_1 dt_2 \Big] + o_p(1) \\
			= & n^{-1} \sum\limits_{i=1}^n {e^{+}}^T E \Big( \iint K_{h_1, h_2} (t_1-t, t_2-t) V_i(t_2, t) \left[V_i(t_1, t)^T \phi_0(t) - V_i(t_2, t)^T \phi(t) \right.\Big.\\
			&\left.\Big. + O_p\{(t_1-t)^2\} \right] \eta(t_1, t_2) dt_1 dt_2 \Big).
		\end{align*}
		Let $t_1=h_1z_1+t, t_2=h_2z_2+t,$ then by conditions (A2)-(A4) and Taylor expansion, we have $E \left[ {e^{+}}^T U_n\{ \phi(t), \hat{\theta}(t) \} \right] = E \{ Z(t)Z(t)^T \} \eta(t, t) \{ \gamma_0(t) - \gamma(t) \} + o(1).$ By the law of large numbers,
		${e^{+}}^T U_{n} \{ \phi(t), \hat{\theta}(t) \} \overset{p}\rightarrow u \{ \gamma(t) \},$ as $n \to \infty,$ where $u \{ \gamma(t) \} = E \{ Z(t)Z(t)^T \} \eta(t, t) \times \newline \left\{ \gamma_0(t) - \gamma(t) \right\}.$ Under condition (A2), $\gamma_0(t)$ is the unique solution to $u \{ \gamma(t) \} = 0.$ $\hat{\gamma}(t)$ solves the estimation equation ${e^{+}}^T U_{n}\{ \phi(t), \hat{\theta}(t) \} = 0.$ By \citet{ag1982}, it follows that for any fixed $t,$ $\hat{\gamma}(t) \stackrel{p}{\to} \gamma_0(t)$. 
		
		We next show the asymptotic normality of $\hat{\gamma}(t).$ Using $\mathcal{P}_n$ and $\mathcal{P}$ to denote the empirical measure and true probability measure respectively, we have
		\begin{align}
			& (nh_1h_2)^{1/2} {e^{+}}^T U_{n} \{ \phi (t), \hat{\theta} (t) \} \notag\\
			= & (nh_1h_2)^{1/2} (\mathcal{P}_n - \mathcal{P}) \left[ \iint K_{h_1, h_2} (t_1 - t, t_2 - t) {e^{+}}^T V_i (t_2, t) \{ Y_i (t_1) - W_i (t_1, t)^T \hat{\theta} (t) -  \right.\notag\\
			& \bigg.- V_i (t_2, t)^T \phi (t) \} dN^*_i(t_1, t_2)\bigg] \notag\\
			& + (nh_1h_2)^{1/2} E \left[ \iint K_{h_1, h_2} (t_1 - t, t_2 - t) {e^{+}}^T V_i (t_2, t) \{ Y_i (t_1) - W_i (t_1, t)^T \hat{\theta} (t) \right. \notag\\
			& \bigg. -  V_i (t_2, t)^T \phi (t) dN^*_i(t_1, t_2) \bigg] \notag\\
			= & \mathrm{\uppercase\expandafter{\romannumeral1}} + \mathrm{\uppercase\expandafter{\romannumeral2}}. \tag{S.14}
			\label{S.14}
		\end{align}
		For the term $\mathrm{\uppercase\expandafter{\romannumeral1}}$ and any fixed time point $t,$ we consider the class of functions
		\begin{align*}
			& \left\{ (h_1 h_2)^{1/2} \iint K_{h_1, h_2} (t_1-t, t_2-t) {e^{+}}^T V(t_2, t) \left\{ Y(t_1) - W(t_1, t)^T \theta (t) - V(t_2, t)^T \phi (t) \right\}\right. \\
			& \bigg.\times dN^*(t_1, t_2): |\theta (t) - \theta_0 (t)| < \epsilon_1, | \phi (t) - \phi_0 (t) | < \epsilon_2 \bigg\}
		\end{align*}
		for given constants $\epsilon_1$ and $\epsilon_2.$ Similarly the proof in Theorem $1$, we obtain that for $\mid \theta (t) - \theta_0 (t) \mid < M_1 (nh)^{-1/2}, \mid \phi (t) - \phi_0 (t) \mid < M_2 (nh_1h_2)^{-1/2}$ where $M_1$ and $M_2$ are some constants, the first term $\mathrm{\uppercase\expandafter{\romannumeral1}}$ in (\ref{S.14}) is equal to
		\begin{equation*}
			\begin{split}
				& (nh_1h_2)^{1/2} (\mathcal{P}_n - \mathcal{P}) \left[ \iint K_{h_1, h_2} (t_1 - t, t_2 - t) {e^{+}}^T V_i (t_2, t) \right. \\
				& \bigg. \times \left\{ Y_i (t_1) - W_i (t_1, t)^T \theta_0 (t) -  V_i (t_2, t)^T \phi_0 (t) \right\} dN^*_i (t_1, t_2) \bigg] + o_p(1)\\
				= & (nh_1h_2)^{1/2} {e^{+}}^T \left( U_n \{ \phi_0(t), \theta_0(t) \} - E \left[ U_n \{ \phi_0 (t), \theta_0 (t) \} \right] \right) + o_p(1).
			\end{split}
		\end{equation*}
		For the second term $\mathrm{\uppercase\expandafter{\romannumeral2}}$ on the right-hand side of equation (\ref{S.14}), we have
		\begin{align*}
			& \mathrm{\uppercase\expandafter{\romannumeral2}} \\
			= & (nh_1h_2)^{1/2} \iint K_{h_1, h_2}(t_1 - t, t_2 - t) E \left[ {e^{+}}^T V(t_2, t) \{ Y(t_1) - W(t_1, t)^T \hat{\theta}(t) -  V(t_2, t)^T \phi(t) \} \right]\\
			& \times \eta(t_1, t_2) dt_1dt_2 \\
			= & (nh_1h_2)^{1/2} E \bigg[ \iint K_{h_1, h_2}(t_1-t, t_2-t) {e^{+}}^T V(t_2, t) \left\{ Y(t_1) - W(t_1, t)^T \theta_0(t) \right.\bigg. \\
			& \bigg.\left.- V(t_2, t)^T \phi_0(t)\right\} \eta(t_1, t_2) dt_1dt_2 \bigg] \\
			& - (nh_1h_2)^{1/2} E\left[ \iint K_{h_1, h_2}(t_1-t, t_2-t) {e^{+}}^T V(t_2, t) W(t_1, t)^T \{\hat{\theta}(t) - \theta_0(t)\} \eta(t_1, t_2) dt_1dt_2 \right] \\
			& - (nh_1h_2)^{1/2} E\left[ \iint K_{h_1, h_2}(t_1-t, t_2-t) {e^{+}}^T V(t_2, t)V(t_2, t)^T \{\phi(t) - \phi_0(t)\} \eta(t_1, t_2) dt_1dt_2 \right] \\
			= & I_1 - I_2 - I_3.
		\end{align*}
		For $I_1,$ let 
		\begin{equation*}
			\begin{split}
				\pi_i(t) = & (h_1h_2)^{1/2} \iint K_{h_1, h_2}(t_1-t, t_2-t) {e^{+}}^T V_i(t_2, t) \left\{ Y_i(t_1) - W_i(t_1, t)^T \theta_0(t) \right.\\
				& \left. - V_i(t_2, t)^T \phi_0(t) \right\} dN^*_i(t_1, t_2),
			\end{split}
		\end{equation*}
		$\pi_i(t)$s are i.i.d. and $(nh_1h_2)^{1/2} {e^{+}}^T U_n\{ \phi_0(t), \theta_0(t) \} = n^{-1/2} \sum_{i=1}^{n} \pi_i(t),$ then we have
		\begin{align*}
			I_1 = & \sqrt{n} E \left\{ \pi(t) \right\} \\
			= & \sqrt{n} E \left[ E\left\{ \pi(t) \mid X(t), Z(s), N^*(t, s), t, s \in [0, 1] \right\} \right] \\
			= & (nh_1h_2)^{1/2} E \bigg[ \iint K_{h_1, h_2} (t_1-t, t_2-t) {e^{+}}^T V(t_2, t) \bigg.\\
			& \bigg. \times E \left\{Y(t_1) - W(t_1, t)^T \theta_0(t) - V(t_2, t)^T \phi_0 (t) \mid X(t), Z(s), t, s \in [0, 1] \right\} \eta (t_1, t_2) dt_1dt_2 \bigg]\\
			= & (nh_1h_2)^{1/2} \iint K_{h_1, h_2} (t_1-t, t_2-t) E \bigg[ {e^{+}}^T \left\{V(t_2, t) V(t_1, t)^T - V(t_2, t) V(t_2, t)^T \right\} \phi_0 (t) \bigg.\\
			& \bigg. + O_p\{(t_1-t)^2\} \bigg] \eta (t_1, t_2) dt_1dt_2, 
		\end{align*}
		let $t_1 = h_1z_1+t, t_2 = h_2z_2+t,$ similar to the proof of Theorem $1$ we have $I_1 = o(1).$ By the proof of Theorem $2$, conditions (A4) and (A8), we have $I_2 = o(1).$ 
		
		\noindent For $I_3,$ we have
		\begin{align*}
			I_3 = & (nh_1h_2)^{1/2} \iint K_{h_1, h_2}(t_1-t, t_2-t) {e^{+}}^T E \{ V(t_2, t) V(t_2, t)^T \} \{ \phi(t) - \phi_0(t) \} \eta(t_1, t_2) dt_1dt_2 \\
			= & (nh_1h_2)^{1/2} \iint K(z_1, z_2) {e^{+}}^T E \{ V(h_2z_2+t, t) V(h_2z_2+t, t)^T \} \eta(h_1z_1+t, h_2z_2+t) dz_1dz_2 \\
			& \times \{ \phi(t) - \phi_0(t) \} \\
			= & (nh_1h_2)^{1/2} E \{ Z(t) Z(t)^T \} \eta(t, t) \{ \gamma(t) - \gamma_0(t) \} + o(1).
		\end{align*}
		Consequently,
		\begin{equation*}
			\begin{split}
				& (nh_1h_2)^{1/2} {e^{+}}^T \left( U_{n} \{ \phi_0(t), \theta_0(t) \} - E \left[ U_{n} \{ \phi_0(t), \theta_0(t) \} \right] \right)\\
				= & (nh_1h_2)^{1/2} E \{Z(t) Z(t)^T\} \eta(t, t) \{ \hat{\gamma}(t) - \gamma_0(t) \} + o_p(1) .
			\end{split}
		\end{equation*}
		Now we show that $(nh_1h_2)^{1/2} {e^{+}}^T U_{n} \{ \phi_0(t), \theta_0(t) \}$ follows the central limit theorem. The variance can be computed using $\pi_i.$
		\begin{equation*}
			\begin{split}
				\mbox{var}\left\{ \pi(t) \right\} = & E \left[ \mathrm{var} \left\{ \pi(t) \mid X(t), Z(s), N^*(t, s), t, s \in [0, 1] \right\} \right] \\
				& + \mathrm{var} \left[ E \left\{ \pi(t) \mid X(t), Z(s), N^*(t, s), t, s \in [0, 1] \right\} \right] \\
				= & J_1 + J_2, 
			\end{split}
		\end{equation*}
		where\begin{align*}
			J_1 = & h_1 h_2 E \bigg[ \iiiint K_{h_1, h_2}(t_1-t, t_2-t) K_{h_1, h_2}(s_1-t, s_2-t) {e^{+}}^T V(t_2, t) V( s_2, t)^T {e^{+}} \bigg.\\
			& \times \bigg. E \left\{ Y(t_1) Y(s_1) \mid X(t), Z(s), N^*(t, s), t, s \in [0, 1] \right\} dN^*(t_1, t_2) dN^*(s_1, s_2) \bigg]\\
			& - h_1 h_2 E \bigg[ \iiiint K_{h_1, h_2} (t_1-t, t_2-t) K_{h_1, h_2} (s_1-t, s_2-t) {e^{+}}^T V(t_2, t) V( s_2, t)^T e^{+} \bigg.\\
			& \times \bigg. E \left\{Y(t_1) \mid X(t), Z(s), N^*(t, s), t, s \in [0, 1]\right\} \bigg. \\
			& \times \bigg.E \left\{Y(s_1) \mid X(t), Z(s), N^*(t, s), t, s \in [0, 1]\right\} dN^*(t_1, t_2) dN^*(s_1, s_2)\bigg]\\
			= & \iint K(z_1, z_2)^2 dz_1 dz_2 E \left[Z(t) Z(t)^T \sigma \{t, X(t), Z(t)\}^2\right] \eta(t, t) + O(h_1^2 + h_2^2), 
		\end{align*}
		and\begin{align*}
			J_2 = & h_1 h_2 \mbox{var} \bigg[ \iint K_{h_1, h_2} (t_1-t, t_2-t) {e^{+}}^T V(t_2, t) \left\{ Z(t_1)^T \gamma_0(t_1) - V(t_2, t)^T \phi_0(t) \right\} \bigg. \\
			& \bigg. \times dN^*(t_1, t_2) \bigg] \\
			= & h_1 h_2 E \bigg[ \iiiint K_{h_1, h_2} (t_1-t, t_2-t) K_{h_1, h_2} (s_1-t, s_2-t) {e^{+}}^T V(t_2, t) \bigg. \\
			& \bigg. \times \left\{ V(t_1, t) - V(t_2, t) \right\}^T \phi_0(t) \phi_0(t)^T \left\{ V(s_1, t) - V(s_2, t) \right\} V( s_2, t)^T e^{+}  \bigg. \\
			& \bigg. \times dN^*(t_1, t_2) dN^*(s_1, s_2) \bigg]\\
			& - h_1 h_2 \bigg( \iint K_{h_1, h_2}(t_1-t, t_2-t) E \left[ {e^{+}}^T V(t_2, t) \{ V(t_1, t) - V(t_2, t) \}^T \right] \phi_0(t) \bigg. \\
			& \bigg. \times dN^*(t_1, t_2) \bigg)^{\otimes 2} \\
			= & O(h_1^2 + h_2^2).
		\end{align*}
		Thus
		\begin{equation*}
			\begin{split}
				\mbox{var}\left\{ \pi(t) \right\} & = \iint K(z_1, z_2)^2 dz_1 dz_2 E \left[Z(t) Z(t)^T \sigma \{t, X(t), Z(t)\}^2\right] \eta(t, t) + O(h_1^2 + h_2^2) \\
				& = \Sigma^{+} (t) + O(h_1^2 + h_2^2).
			\end{split}
		\end{equation*}
		
		To prove the asymptotic normality, we verify the Lyapunov condition. Note that 
		$$(nh_1h_2)^{1/2} {e^{+}}^T U_{n} \left\{ \phi_0(t), \theta_0(t) \right\} = n^{-1/2} \sum\limits_{i=1}^n \pi_i(t) = \sum\limits_{i=1}^n n^{1/2} n^{-1} \pi_i(t),$$ 
		then similar to the calculation of variance,
		\begin{align*}
			\sum\limits_{i=1}^n E \left[\mid n^{1/2} n^{-1} \pi_i(t) - E\{ n^{1/2} n^{-1} \pi_i(t) \}\mid^3\right] &= n O\{(nh_1h_2)^{3/2} n^{-3} (h_1h_2)^{-2}\} \\
			&= O\{(nh_1h_2)^{-1/2}\}.
		\end{align*}
		Consequently, by condition (A4) we have
		$$(nh_1h_2)^{1/2} {e^{+}}^T \left( U_{n} \{ \phi_0(t), \theta_0(t) \} - E \left[ U_{n} \{ \phi_0(t), \theta_0(t) \} \right] \right) \xrightarrow{d} N \{ 0, \Sigma^{+}(t) \}.$$
		Then we have
		$$\sqrt{nh_1h_2} \{ \hat{\gamma}(t) - \gamma_0(t) \} \stackrel{d}{\to} N \{ 0, A^{+}(t)^{-1} \Sigma^{+}(t) A^{+}(t)^{-1} \}, $$
		where \begin{equation*}
			\begin{split}
				A^{+}(t) & = E \{ Z(t) Z(t)^T \} \eta(t, t)\\
				\mbox{and} \quad \Sigma^{+}(t) & =  \big\{ \iint K(z_1, z_2)^2 dz_1 dz_2 \big\}  E\left[ Z(t) Z(t)^T \sigma \{ t, X(t), Z(t) \}^2\right] \eta(t, t).
			\end{split}
		\end{equation*} 
		Therefore the conclusion of Theorem $4$ holds.
	\end{proof}

	\newpage
	\section{Additional simulations}
	
	In this section, we present simulation results for setting (i) $\beta(t)=3(t-0.4)^2$ and $\gamma(t)=\sin(2 \pi t)$ (Table \ref{tabs1}) by using two-step approach with a varying-coefficient model method at the first step and kernel weighting at the second step. We further show results for setting (ii) $\beta(t) = 0.4t + 0.5$ and $\gamma(t) = \sqrt{t}$ using the proposed three methods (Table \ref{tabs2}). 
	
	\begin{table}[!htpb]
		\renewcommand\thetable{S.1}
		\caption{\label{tabs1}1000 simulation results for $\beta(t) = 3(t-0.4)^2$ and  $\gamma(t) = \mbox{sin} (2\pi t)$}
		\centering
		\scalebox{0.6}{
			\begin{threeparttable}
				\begin{tabular}{llrrrrrrrrrrrrrr}
					\hline  
					\hline  
					&&\multicolumn{4}{c}{$t=0.3$}&&\multicolumn{4}{c}{$t=0.6$}&&\multicolumn{4}{c}{$t=0.9$}\\
					\cmidrule{3-6}\cmidrule{8-11}\cmidrule{13-16}
					BD&NP-F&Bias&SD&SE&CP&&Bias&SD&SE&CP&&Bias&SD&SE&CP\\
					\hline 
					{Results for $n = 400$}&&\multicolumn{14}{c}{Two-step \ (VCM+KW)}\\
					$h = n^{-0.6}, h_1 = h_2 = n^{-0.5}$&$\beta (t)$&$0.003$&$0.157$&$0.144$&$92.2$&&$0.005$&$0.124$&$0.122$&$92.9$&&$-0.001$&$0.133$&$0.122$&$92.5$\\
					$h = n^{-0.7}, h_1 = h_2 = n^{-0.5}$&$\beta (t)$&$0.004$&$0.198$&$0.186$&$92.2$&&$-0.001$&$0.168$&$0.155$&$91.3$&&$0.006$&$0.168$&$0.156$&$92.6$\\
					auto&$\beta (t)$&$-0.007$&$0.154$&$0.145$&$91.5$&&$0.003$&$0.130$&$0.126$&$92.9$&&$-0.003$&$0.127$&$0.125$&$93.5$\\
					$h = n^{-0.6}, h_1 = h_2 = n^{-0.5}$&$\gamma (t)$&$-0.045$&$0.153$&$0.144$&$90.3$&&$0.033$&$0.155$&$0.140$&$88.8$&&$0.034$&$0.147$&$0.134$&$89.7$\\
					$h = n^{-0.7}, h_1 = h_2 = n^{-0.5}$&$\gamma (t)$&$-0.045$&$0.163$&$0.151$&$89.7$&&$0.029$&$0.166$&$0.148$&$89.1$&&$0.028$&$0.148$&$0.142$&$91.0$\\
					auto&$\gamma (t)$&$-0.039$&$0.153$&$0.141$&$90.6$&&$0.035$&$0.148$&$0.141$&$91.4$&&$0.032$&$0.133$&$0.128$&$92.0$\\
					&\\
					{Results for $n = 900$}&&\multicolumn{14}{c}{Two-step \ (VCM+KW)}\\
					$h = n^{-0.6}, h_1 = h_2 = n^{-0.5}$&$\beta (t)$&$0.001$&$0.120$&$0.119$&$94.6$&&$-0.003$&$0.101$&$0.100$&$93.8$&&$0.001$&$0.102$&$0.100$&$93.4$\\
					$h = n^{-0.7}, h_1 = h_2 = n^{-0.5}$&$\beta (t)$&$0.007$&$0.171$&$0.161$&$92.1$&&$0.000$&$0.140$&$0.137$&$94.1$&&$0.002$&$0.147$&$0.136$&$91.9$\\
					auto&$\beta (t)$&$-0.002$&$0.134$&$0.125$&$92.8$&&$0.000$&$0.104$&$0.100$&$93.3$&&$-0.001$&$0.103$&$0.100$&$92.6$\\
					$h = n^{-0.6}, h_1 = h_2 = n^{-0.5}$&$\gamma (t)$&$-0.024$&$0.140$&$0.131$&$91.0$&&$0.018$&$0.135$&$0.130$&$92.1$&&$0.028$&$0.127$&$0.122$&$91.6$\\
					$h = n^{-0.7}, h_1 = h_2 = n^{-0.5}$&$\gamma (t)$&$-0.028$&$0.147$&$0.156$&$90.8$&&$0.013$&$0.138$&$0.135$&$92.7$&&$0.016$&$0.130$&$0.139$&$92.5$\\
					auto&$\gamma (t)$&$-0.036$&$0.141$&$0.129$&$90.2$&&$0.014$&$0.135$&$0.129$&$92.7$&&$0.021$&$0.116$&$0.123$&$92.1$\\
					\hline 
				\end{tabular}
				\begin{tablenotes}
					\footnotesize
					\item Note: ``BD" represents the bandwidths, where $h$ represents the bandwidth in the varying coefficient model (VCM) approach, $h_1$ and $h_2$ represent the bandwidths in the kernel weighting (KW) approach. ``NP-F" represents the non-parametric function. ``Bias" is the absolute bias. ``SD" is the sample standard deviation. ``SE" is the average of the standard error and ``CP" is the point-wise $95\%$ coverage probability. ``auto" represents automatic bandwidth selection.
				\end{tablenotes}
		\end{threeparttable}}
	\end{table}
	\begin{table}[!htpb]
		\renewcommand\thetable{S.2}
		\caption{\label{tabs2}1000 simulation results for $\beta(t) = 0.4t+0.5$ and  $\gamma(t) = \sqrt{t}$}
		\centering
		\scalebox{0.6}{
			\begin{threeparttable}
				\begin{tabular}{llrrrrrrrrrrrrrr}
					\hline  
					\hline  
					&&\multicolumn{4}{c}{$t=0.3$}&&\multicolumn{4}{c}{$t=0.6$}&&\multicolumn{4}{c}{$t=0.9$}\\
					\cmidrule{3-6}\cmidrule{8-11}\cmidrule{13-16}
					BD&NP-F&Bias&SD&SE&CP&&Bias&SD&SE&CP&&Bias&SD&SE&CP\\
					\hline 
					{Results for $n = 400$}&&\multicolumn{14}{c}{Two-step \ (Centering+KW)}\\
					$h = n^{-0.6}, h_1 = h_2 = n^{-0.5}$&$\beta (t)$&$0.003$&$0.130$&$0.120$&$91.4$&&$-0.003$&$0.138$&$0.133$&$92.6$&&$-0.004$&$0.150$&$0.145$&$93.1$\\
					$h = n^{-0.7}, h_1 = h_2 = n^{-0.5}$&$\beta (t)$&$0.010$&$0.164$&$0.155$&$92.0$&&$0.000$&$0.177$&$0.172$&$92.6$&&$0.004$&$0.200$&$0.186$&$91.8$\\
					auto&$\beta (t)$&$-0.003$&$0.142$&$0.133$&$91.9$&&$0.005$&$0.134$&$0.132$&$93.1$&&$-0.001$&$0.164$&$0.154$&$92.6$\\
					$h = n^{-0.6}, h_1 = h_2 = n^{-0.5}$&$\gamma (t)$&$-0.025$&$0.147$&$0.138$&$91.6$&&$-0.031$&$0.152$&$0.140$&$88.9$&&$-0.035$&$0.149$&$0.135$&$89.6$\\
					$h = n^{-0.7}, h_1 = h_2 = n^{-0.5}$&$\gamma (t)$&$-0.026$&$0.158$&$0.150$&$92.5$&&$-0.027$&$0.156$&$0.149$&$91.0$&&$-0.029$&$0.152$&$0.144$&$90.8$\\
					auto&$\gamma (t)$&$-0.024$&$0.144$&$0.133$&$90.4$&&$-0.029$&$0.151$&$0.143$&$91.0$&&$-0.029$&$0.144$&$0.136$&$91.2$\\
					&&\multicolumn{14}{c}{Two-step \ (VCM+KW)}\\
					$h = n^{-0.6}, h_1 = h_2 = n^{-0.5}$&$\beta (t)$&$-0.002$&$0.131$&$0.119$&$91.8$&&$-0.011$&$0.132$&$0.132$&$94.2$&&$-0.004$&$0.147$&$0.144$&$93.1$\\
					$h = n^{-0.7}, h_1 = h_2 = n^{-0.5}$&$\beta (t)$&$-0.002$&$0.167$&$0.154$&$91.6$&&$-0.002$&$0.187$&$0.169$&$91.2$&&$0.001$&$0.199$&$0.188$&$92.7$\\
					auto&$\beta (t)$&$0.004$&$0.130$&$0.128$&$93.7$&&$0.012$&$0.144$&$0.134$&$92.0$&&$0.004$&$0.151$&$0.144$&$91.9$\\
					$h = n^{-0.6}, h_1 = h_2 = n^{-0.5}$&$\gamma (t)$&$-0.029$&$0.152$&$0.136$&$89.9$&&$-0.028$&$0.149$&$0.141$&$92.3$&&$-0.033$&$0.145$&$0.134$&$90.5$\\
					$h = n^{-0.7}, h_1 = h_2 = n^{-0.5}$&$\gamma (t)$&$-0.026$&$0.159$&$0.143$&$89.6$&&$-0.032$&$0.157$&$0.153$&$91.0$&&$-0.039$&$0.154$&$0.150$&$91.6$\\
					auto&$\gamma (t)$&$-0.025$&$0.142$&$0.133$&$91.1$&&$-0.030$&$0.151$&$0.137$&$90.4$&&$-0.043$&$0.140$&$0.129$&$90.4$\\
					&&\multicolumn{14}{c}{One-step \ (KW)}\\
					$h_1 = h_2 = n^{-0.45}$&$\beta (t)$&$0.001$&$0.124$&$0.114$&$91.8$&&$-0.004$&$0.118$&$0.115$&$91.8$&&$-0.005$&$0.116$&$0.119$&$93.9$\\
					$h_1 = h_2 = n^{-0.5}$&$\beta (t)$&$0.004$&$0.146$&$0.136$&$90.6$&&$-0.005$&$0.148$&$0.135$&$90.3$&&$-0.002$&$0.150$&$0.138$&$91.5$\\
					auto&$\beta (t)$&$-0.001$&$0.101$&$0.096$&$93.1$&&$0.006$&$0.125$&$0.114$&$91.7$&&$0.000$&$0.111$&$0.103$&$92.3$\\
					$h_1 = h_2 = n^{-0.45}$&$\gamma (t)$&$-0.025$&$0.122$&$0.115$&$92.6$&&$-0.024$&$0.124$&$0.115$&$91.4$&&$-0.030$&$0.120$&$0.111$&$90.7$\\
					$h_1 = h_2 = n^{-0.5}$&$\gamma (t)$&$-0.009$&$0.148$&$0.137$&$91.0$&&$-0.020$&$0.148$&$0.135$&$90.2$&&$-0.018$&$0.151$&$0.135$&$91.6$\\
					auto&$\gamma (t)$&$-0.026$&$0.101$&$0.096$&$91.7$&&$-0.026$&$0.122$&$0.113$&$91.0$&&$-0.040$&$0.102$&$0.098$&$91.2$\\
					&\\
					{Results for $n = 900$}&&\multicolumn{14}{c}{Two-step \ (Centering+KW)}\\
					$h = n^{-0.6}, h_1 = h_2 = n^{-0.5}$&$\beta (t)$&$-0.001$&$0.099$&$0.098$&$93.7$&&$-0.004$&$0.115$&$0.110$&$92.8$&&$0.007$&$0.118$&$0.119$&$94.2$\\
					$h = n^{-0.7}, h_1 = h_2 = n^{-0.5}$&$\beta (t)$&$0.000$&$0.143$&$0.134$&$92.9$&&$0.001$&$0.157$&$0.147$&$90.6$&&$0.000$&$0.166$&$0.162$&$93.4$\\
					auto&$\beta (t)$&$0.000$&$0.112$&$0.110$&$93.9$&&$0.002$&$0.114$&$0.109$&$92.7$&&$0.001$&$0.139$&$0.134$&$92.7$\\
					$h = n^{-0.6}, h_1 = h_2 = n^{-0.5}$&$\gamma (t)$&$-0.015$&$0.137$&$0.126$&$90.4$&&$-0.016$&$0.134$&$0.127$&$92.3$&&$-0.015$&$0.126$&$0.124$&$92.3$\\
					$h = n^{-0.7}, h_1 = h_2 = n^{-0.5}$&$\gamma (t)$&$-0.010$&$0.140$&$0.143$&$91.6$&&$-0.021$&$0.142$&$0.143$&$92.5$&&$-0.025$&$0.140$&$0.177$&$92.6$\\
					auto&$\gamma (t)$&$-0.014$&$0.132$&$0.128$&$91.3$&&$-0.014$&$0.134$&$0.127$&$92.1$&&$-0.024$&$0.133$&$0.129$&$91.2$\\
					&&\multicolumn{14}{c}{Two-step \ (VCM+KW)}\\
					$h = n^{-0.6}, h_1 = h_2 = n^{-0.5}$&$\beta (t)$&$-0.002$&$0.104$&$0.099$&$93.0$&&$0.004$&$0.113$&$0.109$&$93.9$&&$0.000$&$0.123$&$0.119$&$92.8$\\
					$h = n^{-0.7}, h_1 = h_2 = n^{-0.5}$&$\beta (t)$&$0.004$&$0.143$&$0.133$&$91.4$&&$0.000$&$0.157$&$0.148$&$92.3$&&$0.004$&$0.166$&$0.160$&$93.4$\\
					auto&$\beta (t)$&$-0.001$&$0.095$&$0.098$&$94.9$&&$0.006$&$0.116$&$0.113$&$93.6$&&$0.004$&$0.123$&$0.119$&$92.9$\\
					$h = n^{-0.6}, h_1 = h_2 = n^{-0.5}$&$\gamma (t)$&$-0.009$&$0.130$&$0.127$&$92.9$&&$-0.015$&$0.132$&$0.127$&$92.6$&&$-0.022$&$0.129$&$0.124$&$91.9$\\
					$h = n^{-0.7}, h_1 = h_2 = n^{-0.5}$&$\gamma (t)$&$-0.016$&$0.141$&$0.132$&$91.6$&&$-0.026$&$0.138$&$0.138$&$92.2$&&$-0.025$&$0.136$&$0.155$&$91.6$\\
					auto&$\gamma (t)$&$-0.013$&$0.124$&$0.119$&$92.6$&&$-0.020$&$0.125$&$0.119$&$92.2$&&$-0.031$&$0.131$&$0.123$&$90.5$\\
					&&\multicolumn{14}{c}{One-step \ (KW)}\\
					$h_1 = h_2 = n^{-0.45}$&$\beta (t)$&$0.002$&$0.102$&$0.098$&$94.2$&&$-0.003$&$0.107$&$0.099$&$92.9$&&$0.003$&$0.106$&$0.100$&$92.6$\\
					$h_1 = h_2 = n^{-0.5}$&$\beta (t)$&$0.000$&$0.136$&$0.125$&$90.4$&&$0.002$&$0.135$&$0.126$&$91.7$&&$-0.001$&$0.139$&$0.128$&$92.6$\\
					auto&$\beta (t)$&$-0.001$&$0.086$&$0.082$&$93.1$&&$0.001$&$0.085$&$0.083$&$93.3$&&$0.001$&$0.100$&$0.098$&$93.4$\\
					$h_1 = h_2 = n^{-0.45}$&$\gamma (t)$&$-0.011$&$0.104$&$0.099$&$90.9$&&$-0.017$&$0.105$&$0.098$&$92.0$&&$-0.018$&$0.099$&$0.094$&$91.8$\\
					$h_1 = h_2 = n^{-0.5}$&$\gamma (t)$&$-0.006$&$0.131$&$0.125$&$92.3$&&$-0.015$&$0.133$&$0.126$&$92.8$&&$-0.012$&$0.134$&$0.121$&$90.8$\\
					auto&$\gamma (t)$&$-0.021$&$0.086$&$0.081$&$91.7$&&$-0.026$&$0.088$&$0.083$&$91.1$&&$-0.025$&$0.095$&$0.093$&$93.1$\\
					\hline 
				\end{tabular}
				\begin{tablenotes}
					\footnotesize
					\item Note: ``BD" represents the bandwidths, where $h$ represents the bandwidth in the centering and varying coefficient model (VCM) approaches, $h_1$ and $h_2$ represent the bandwidths in the kernel weighting (KW) approach. ``NP-F" represents the non-parametric function. ``Bias" is the absolute bias. ``SD" is the sample standard deviation. ``SE" is the average of the standard error and ``CP" is the point-wise $95\%$ coverage probability. ``auto" represents automatic bandwidth selection.
				\end{tablenotes}
		\end{threeparttable}}
	\end{table}
	
\end{document}